\documentclass{aa}
\usepackage{graphicx}
\def\hmpc{~h$^{-1}$ Mpc~}

\def\parn{\par\noindent}
\def\ltsim{\raise 2pt \hbox {$<$} \kern-1.1em \lower 4pt \hbox {$\sim$}}
\def\ltapprox{\raise 2pt \hbox {$<$} \kern-1.1em \lower 5pt \hbox {$\approx
$}}
\def\simeq{\raise 2pt \hbox{$\sim$} \kern-1.1em \lower 4pt \hbox {$-$}}

\def\gtsim{\raise 2pt \hbox {$>$} \kern-1.1em \lower 4pt \hbox {$\sim$}}
\def\gtapprox{\raise 2pt \hbox {$>$} \kern-1.1em \lower 5pt \hbox {$\approx
$}}

\begin{document}
%
%
%
\title{AGN and starburst radio activity in the A3558 cluster complex}
%
%
\author
{
S.~Giacintucci\inst{1} \and
T.~Venturi\inst{2} \and
S.~Bardelli\inst{1} \and
D.~Dallacasa\inst{2,}\inst{3}\and 
E.~Zucca\inst{1}
}

\institute
{
INAF - Osservatorio Astronomico di Bologna, 
via Ranzani 1, I--40127 Bologna, Italy
\and
Istituto di Radioastronomia del CNR, via Gobetti 101, I-40129, Bologna, Italy 
\and
Dipartimento di Astronomia, Universit\`a di Bologna,
via Ranzani 1, I--40127, Bologna, Italy
}
%
\date{Received 00 - 00 - 0000; accepted 00 - 00 - 0000}
%
\titlerunning{Radio activity in A3558}
\authorrunning{S. Giacintucci et al.}
\abstract{
In this paper we present Very Large Array (VLA) 1.4 GHz (21 cm) 
observations of the region between the centres of A3558 and A3562, in the major
cluster merger complex of the Shapley Concentration.
Our final catalogue includes a total of 174 radio sources above the 
flux density limit of 0.25 mJy b$^{-1}$. By cross-correlation with optical 
and spectroscopic catalogues we found 33 optical counterparts 
belonging to the Shapley Concentration.
\\
We investigated the effects of cluster merger on the radio emission 
properties of the galaxy population by means of the radio source 
counts and the radio luminosity functions (RLF). We found that the radio 
source counts are consistent with the field source counts.
The RLF of elliptical and S0 galaxies in the region surveyed here, 
is consistent with the ``universal'' RLF  for early--type galaxies.
This result suggests that the deficit in radio galaxies found in our
previous works over the whole A3558 chain, is entirely due to the
cluster A3558.
A population of faint radio 
galaxies (logP$_{1.4~GHz}$(W Hz$^{-1}$) \ltsim  22) is also found.
Half of these objects are also blue, suggesting that
starburst is the main mechanism driving the radio emission.
Finally, we detected 14 spiral radio galaxies, whose ratio between 
radio and optical emission is similar to those found in galaxies located
in rich and dynamically evolved clusters.
\\
Our results are briefly discussed in the light of the age and stage
of the merger in the A3558 cluster complex.
}
\maketitle

%
%

\section{Introduction}

Evidence is accumulating that the interaction processes between 
clusters of galaxies, known as cluster mergers, may significantly
affect the radio emission characteristics of the cluster galaxy
population. In particular, the evolution of galaxies and the
properties of their nuclear and/or star forming activity are
undoubtely influenced by the interaction with the environment.
In this frame, cluster merging and group accrection seem to play
an important role, but it is not yet completely understood how
the merging environment affects the nuclear and starburst emission
in galaxies. Owen et al. (\cite{owen99}) suggested that merging 
may trigger the radio emission, both in the form of nuclear activity
and starburst phenomena. Burns et al. (\cite{burns94}) interpreted 
the presence of post-starburst galaxies in the X--ray and radio bridge 
connecting the Coma cluster and the NGC 4839 group as consequence 
of a recent merging process between these two structures.
Using numerical simulations, Bekki (\cite{bekki99}) concluded that the tidal
gravitational field of a cluster merger may drive a considerable
transfer of gas to the central regions of galaxies, leading to 
enhanced star formation activity or feeding the central engine of 
active galactic nuclei. Vollmer et al. (\cite{vollmer01b})
suggested that a local burst of star formation could be due to
re-accretion of gas after a ram pressure stripping event, which
is thought to be responsible for the HI deficit of spiral galaxies
in the central regions of nearby clusters with respect to the
field galaxies of the same morphological type and optical size (Gunn
\& Gott, \cite{gunn72}; Bothun \& Dressler, \cite{bothun86}).
On the other hand, Fujita et al. (\cite{fujita99}) and
Balogh et al. (\cite{balogh98}) claimed that gas stripping resulting from
ram pressure is important in preventing gas supply to the central
regions of the galaxies and may suppress star formation. Finally Venturi et al. 
(\cite{venturi00}, hereinafter V2000) found evidence that merging may 
anticorrelate with the radio emission, possibly switching off previously 
existing radio galaxies, or temporarily inhibiting the nuclear radio
activity.
\\
A major problem in our current knowledge of the role of
cluster mergers on the radio emission and in the interpretation of 
the observations is the lack of statistics. Only few clusters
of galaxies were deeply imaged over their whole extent (i.e. A2125,
Dwarakanath \& Owen, 1999; A2255, Miller \& Owen, 2003 (hereinafter MO03); 
A2256, Miller et al., 2003; A2645, Owen et al., 1999;
the Shapley supercluster chains  A3528, A3558 and A3571, 
Venturi et al. 2001, 2000 and 2002 respectively), and the complexity 
of the phenomenon clearly needs a much larger observational support.
Cluster mergers evolve over a timescale of $\sim$ Gyr, while typical
ages of radio sources are of the order of few times 10$^7$ -- 10$^8$
years, therefore it is of crucial importance to identify the
effects of the various stages of a merger event. For this reason
it is essential to carry out deep radio observations 
of a large number of galaxy clusters, at different merger stages.
\\
In order to study the effects of a major cluster merger event after the first 
core--core encounter on the nuclear and starburst radio emission, 
in this paper we present 1.4 GHz Very Large Array (VLA) observations
of the cluster A3562 and the poor groups SC 1329$-$313 and SC 1327$-$312,
at the eastern side of the A3558 cluster complex.
In the following we will refer to the area surveyed as A3558--C. 
This work is part of a larger project which aims
to study the influence of the ongoing merger in the core of the Shapley
Concentration on the radio/optical properties of the cluster galaxies,
both from a statistical point of view and through a detailed analysis
of the physical properties of the extended radio galaxies.
In section \ref{sec:intro} we briefly overview the optical and X--ray
properties of A3558--C, with emphasis on the cluster A3562 and on the 
poor groups SC 1329$-$313 and SC 1327$-$312; 
the observations and data reduction are presented 
in Section \ref{sec:obs}; in Section \ref{sec:sample} we present 
the 1.4 GHz radio sample and the source counts; 
in Section \ref{sec:ident} we deal with the 
optical identifications and in Section \ref{sec:prop} we present the 
general properties of the radio galaxies in this region; the
results of the analysis of the radio luminosity function (RLF) for 
the early-type and late-type galaxy populations are given in 
Section \ref{sec:RLF}; discussion and 
conclusions are given in Section \ref{sec:disc}.

We assume $H_0 = 100$ km s$^{-1}$ Mpc$^{-1}$ and $q_o$=0.5. If we define 
h=H$_o$/100, at the average redshift of the Shapley Concentration 
($z=0.05$) this 
leads to a linear scale of 1 arcsec = 0.67 h$^{-1}$ kpc. 
We will assume $S \propto \nu^{-\alpha}$.

\vskip 1 truecm \noindent
\section{The cluster merger in A3558--C} \label{sec:intro} 

\begin{figure*}
\centering
\includegraphics[angle=0,width=\hsize]{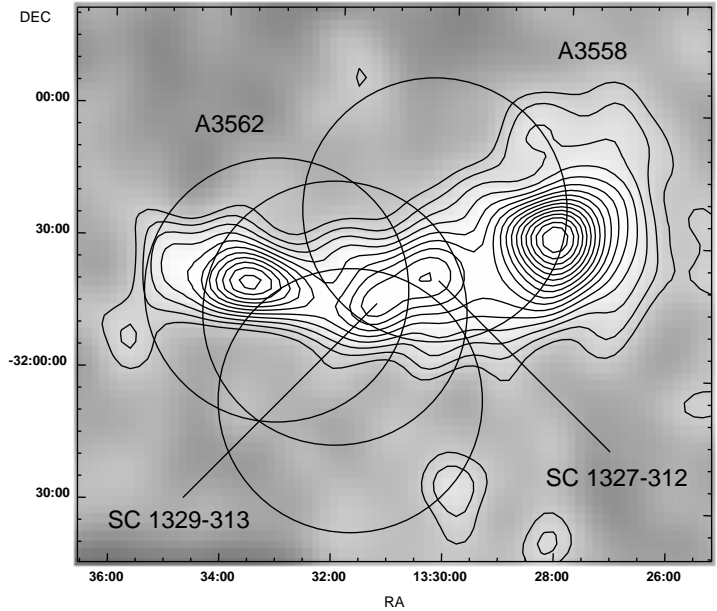}
\caption{Grey scale and isocontours of the galaxy density to $b_J$ = 19.5 
in A3558--C from the COSMOS catalogue. The cluster A3556, located westward of 
A3558, is not included in the figure. The centres of the superposed circles 
correspond to the pointing centres of the observations. The radius of each 
circle is 30 arcmin (i.e. primary beam size of the VLA at 1.4 GHz).} 
\label{fig:A3562}
\end{figure*}

The dynamical centre of the Shapley Supercluster can be
identified with the A3558 cluster complex, a chain of interacting 
clusters located at  a mean redshift $z=0.0483$ and spanning 
$\sim 7.5$\hmpc (projected), almost orthogonal to the line of sight.
The chain is formed by the
three ACO (Abell et al. 1989) clusters A3558, A3562 and A3556, 
and includes the two poor groups SC 1327$-$312 and SC 1329$-$313 
(Figure \ref{fig:A3562}).
%
%
%
%

The A3358 cluster complex has been extensively studied over a wide range of 
bands, from radio wavelengths up to X--ray energies, and 
the physical connection and the merging stage of all clusters
in this region is suggested by observational data 
in the radio (V2000, Venturi et al. 2003, hereinafter V2003), optical 
(Bardelli et al. \cite{bardelli94}) and X--ray bands 
(Bardelli et al. \cite{bardelli96}, Ettori et al. \cite{ettori97}).
\\
The distribution of the hot gas in this region remarkably 
follows the distribution of the optical galaxies, and a detailed substructure 
analysis carried out by Bardelli et al. (\cite{bardelli98a}) revealed that 
the whole structure is characterized by a large number of subcondensations,
further evidence of its dynamical activity.
In addition, the galaxy distribution in the A3558 chain resembles the 
results of the numerical simulations carried out by Roettiger et al. (\cite{roettiger97})
for the case of a collision of two 
clusters. The observational scenario suggests that the whole region
is the result of a major merger with the cluster A3558, seen after the
first core--core encounter (Bardelli et al. \cite{bardelli98b}, \cite{bardelli02}). 
In this frame, all that visible in the chain beyond A3558 itself, 
i.e. A3556, A3562 and the 
two small groups SC 1327$-$312 and SC 1329$-$313, would be the remains of the 
colliding cluster (see Fig. \ref{fig:A3562}).
In the region between A3558 and A3562, where the initial position of the merger 
shock is expected, Bardelli et al. (\cite{bardelli98b}) found an enhanced fraction 
of blue galaxies. 
\\
The whole A3558 cluster complex was surveyed at 22/13 cm with 
the Australia Telescope Compact Array (ATCA, Venturi et al. \cite{venturi97}, 
Venturi et al. \cite{venturi98}, V2000) and a number of interesting results 
emerged from the radio analysis.
In particular, the outskirts of the chain, i.e. A3556 and A3562, contain
the largest number of cluster radio sources and the four extended radio 
galaxies. Moreover, the radio luminosity 
function computed for elliptical galaxies over the whole complex is considerably 
lower that the ``universal'' radio luminosity function for ellipticals derived 
by Ledlow and Owen (\cite{ledlow96}, hereinafter LO96), suggesting a possible 
connection with the cluster merger in this region.
Finally, a cluster radio halo was detected at the centre of A3562 (V2000, 
V2003), as a further evidence of the role of merger 
on the cluster radio emission.
\\
The properties of A3562, SC 1327-312 and SC 1329$-$313 are summarised in 
Table \ref{tab:prop}, where we report:

\parn $-$ columns 2 and 3: J2000 coordinates;

\parn $-$ columns 4 and 5: Bautz-Morgan type and richness;

\parn $-$ columns 6 and 7: mean heliocentric velocity and velocity dispersion;

\parn $-$ column 8 and 9: X--ray temperature and luminosity.

%
%
\begin{table*}
\caption[]{Cluster properties}
\begin{center}
\begin{tabular}{lcccccccc}
\hline\noalign{\smallskip}
Cluster & RA$_{J2000}$ & DEC$_{J2000}$ & B--M & R & $<v>$       & $\sigma_v $ & kT  &
L$_X$ (2--10 keV)  \\
        &              &               &          &   & km s$^{-1}$ & km s$^{-1}$ & keV & 
h$^{-2}$ erg s$^{-1}$ \\
(1) & (2) & (3) & (4) & (5) & (6) & (7) & (8) & (9) \\
\\
\noalign{\smallskip}
\hline\noalign{\smallskip}
SC 1327$-$312 &   13 29 47   &  $-$31 36 29  &  -- & -- & 14844$^{+105}_{-211}$ (a) &
691$^{+158}_{-146}$ (a) & 4.11$^{+0.43}_{-0.36}$ (c) & 3.76$\times 10^{43}$ (c) \\
&&&&&&&&\\
SC 1329$-$313 &   13 31 36   &  $-$31 48 46  &  -- & -- & (T520) 13280$^{+80}_{-95}$ (b) &
482$^{+87}_{-49}$ (b)   & 3.49$^{+0.27}_{-0.24}$ (c) & 2.36$\times 10^{43}$ (c)  \\
             &              &               &     &    & (T496) 14690$^{-101}_{-99}$ (b) &
537$^{+87}_{-32}$ (b)   &                            &                           \\
&&&&&&&&\\
A3562        &   13 33 30   &  $-$31 40 00  &  I  &  2 & 14492$^{+225}_{-286}$ (a) & 
913$^{189}_{-96}$ (a) & 5.13$^{+0.21}_{-0.19}$ (d) & 5.25$\times 10^{43}$ (d) \\
\\
\noalign{\smallskip}
\hline
\end{tabular}
\end{center}
\begin{flushleft}
\vspace{0.5cm}
Notes to Table 1.
The coordinates
of the centre for A3562 are taken from Abell et al. (\cite{abell89}),
the coordinates for SC 1327--312 are from Bardelli et al. (\cite{bardelli96}) and for
SC 1329--313 from Breen et al. (\cite{breen94}). Note that the SC 1329 group has a 
bimodal velocity distribution: the optical clumps associated to the two velocity 
peaks are named T520 and T496 (Bardelli et al., \cite{bardelli98a}).
(a) Bardelli et al. (\cite{bardelli98b}); (b) Bardelli et al. (\cite{bardelli98a});
(c) Bardelli et al. (\cite{bardelli02});
(d) Ettori et al. (\cite{ettori00}). 
\end{flushleft}
\label{tab:prop}
\end{table*}
\medskip 
%
%

\section{Observations and Data Reduction}\label{sec:obs}

The 1.4 GHz (21 cm) observations of A3562 and of the two SC groups 
1327$-$312 and  1329$-$313 were carried out on 10 July 2000 
with the Very Large Array (VLA), operating in the DnC configuration.
The observations were carried out in continuum mode with a 50
MHz bandwidth, for a total duration of 4 hours. We switched
among the four fields every 4.5 minutes. 3C286 was used as primary calibrator, 
and 1316$-$336 as phase calibrator.
The logs of the observations
are given in Table \ref{tab:observations}.

The region of interest was covered with four different pointings;
the resolution is $41.98^{\prime \prime}$ $\times$ 35.13$^{\prime \prime}$
at the declination of the A3558 complex. 

The observations were carried out using the mosaicing 
facility of the VLA. Pointing $\#$ 1 and $\#$ 2 were chosen to be centered 
respectively on the head-tail J1333$-$3141 in the centre of A3562, where 
the radio halo lies, and on the diffuse radio source
J1332$-$3146a in the region between A3562 and the SC 1329$-$313 group (V2000).
The remaining pointings were chosen in order to remove the
sidelobes from the confusing sources at the field edges and to ensure
uniform sensitivity over the whole region between the centre
of A3562 and the SC groups.

%
%
\begin{table*}
\caption[]{Logs of the Observations}
\begin{center}
\begin{tabular}{cccccc}
\hline\noalign{\smallskip}
Field & RA$_{J2000}$ & DEC$_{J2000}$ &Array & Int. Time &  rms (1.4 GHz) \\
\#    &              &               &      &  (h)      &    mJy/b \\
(1)   &   (2)        &   (3)         & (4)  &   (5)     &    (6)   \\
\\
\noalign{\smallskip}
\hline\noalign{\smallskip}
1   & 13 33 32  & $-$31 40 60 & DnC  &  1  & 0.05  \\
2   & 13 32 02  & $-$31 46 60 & DnC  &  1  & 0.05  \\
3   & 13 30 19  & $-$31 22 57 & DnC  &  1  & 0.05  \\
4   & 13 31 43  & $-$32 06 45 & DnC  &  1  & 0.05  \\
\noalign{\smallskip}
\hline
\end{tabular}
\end{center}
\label{tab:observations}
\end{table*}
%
%

The data reduction was carried out using the standard
procedure (calibration, Fourier inversion, clean and restore)
of the NRAO AIPS (Astronomical Image Processing System) package. 
We reduced the data and imaged each field separately. The final
images were then mosaiced (i.e. linearly combined) using the AIPS task LTESS.
The average noise in the individual final images is $\sim$0.50 $\mu$Jy b$^{-1}$ 
in all four fields.
We assume as reliable all detections with flux density peak 
S$_{1.4GHz}\ge$0.25 mJy b$^{-1}$, corresponding to 5$\sigma$ confidence 
level. Such limit implies a limit on radio power of 
logP$_{1.4GHz}$ (W Hz$^{-1}$) = 20.83 at the average distance of  
A3558--C. We note that our detection limit is near the confusion
limit at the frequency and resolution of the images presented here. In
order to evaluate this effect, we cross--checked our detections on a 
pure uniform weight image, where the confusion is considerably lower,
ad considered as reliable detections only those sources passing the
cross-check.

\section{The sample of radio sources}\label{sec:sample}

We detected a total of 174 radio sources at 1.4 GHz above the peak flux density
limit of 0.25 mJy b$^{-1}$. The source list is reported in Table 3,
where we give:

\parn $-$ columns 1, 2 and 3:  name (VLA--) and J2000 position;

\parn $-$ column 4: flux density at 1.4 GHz corrected for the primary beam 
attenuation. The values were derived with task JMFIT for unresolved
or marginally resolved sources, while they were obtained by means 
of TVSTAT for the extended sources;

\parn $-$ column 5: radio morphology. We classified the sources as unres. = 
unresolved and ext. = extended. Moreover we note that D = double and 
HT = head--tail. 
For the double source we give the position of the radio barycentre and for 
the extended sources we give the position of the radio peak.

%
%
\begin{table*}
\label{tab:cat}
\caption[]{Source list and flux density values}
\begin{center}
\begin{tabular}{lcccc}
\hline\noalign{\smallskip}
Name  & RA$_{J2000}$   &  DEC$_{J2000}$ & S$_{1.4GHz}$   & Radio Morphology \\
VLA$-$&                &                &    (mJy)       &                  \\ 
\noalign{\smallskip}
\hline\noalign{\smallskip}
J1328$-$3119   &  13 28 29.35   &  $-$31 19 14.7   &   66.34  &   unres.  \\
J1328$-$3134   &  13 28 31.50   &  $-$31 34 50.4   &  126.80  &   unres.  \\
J1328$-$3115   &  13 28 35.09   &  $-$31 15 22.1   &    1.97  &   unres.  \\
J1329$-$3126   &  13 29 00.02   &  $-$31 26 26.0   &   81.92  &    ext.   \\
J1329$-$3131   &  13 29 04.61   &  $-$31 31 02.1   &   86.80  &   unres.  \\
J1329$-$3112   &  13 29 09.12   &  $-$31 12 56.8   &    5.37  &   unres.  \\
J1329$-$3121   &  13 29 13.32   &  $-$31 21 51.1   &   23.65  &   unres.  \\
J1329$-$3129a  &  13 29 29.43   &  $-$31 29 43.7   &   18.90  &   unres.  \\
J1329$-$3100   &  13 29 31.67   &  $-$31 00 17.0   &    5.48  &   unres.  \\
J1329$-$3116   &  13 29 31.88   &  $-$31 16 51.9   &   25.06  &    ext.   \\
J1329$-$3102   &  13 29 39.15   &  $-$31 02 08.3   &    2.15  &   unres.  \\
J1329$-$3119a  &  13 29 49.32   &  $-$31 19 37.3   &    6.02  &   unres.  \\
J1329$-$3101   &  13 29 49.42   &  $-$31 01 44.6   &    3.03  &   unres.  \\ 
J1329$-$3122   &  13 29 50.76   &  $-$31 22 55.6   &   22.43  &    ext.   \\
J1329$-$3057   &  13 29 51.22   &  $-$30 57 47.2   &   65.38  &   unres.  \\
J1329$-$3056   &  13 29 52.90   &  $-$30 56 00.7   &  123.04  & ext. (D)  \\
J1329$-$3119b  &  13 29 55.02   &  $-$31 19 57.7   &    4.53  &   unres.  \\
J1329$-$3129b  &  13 29 55.32   &  $-$31 29 36.7   &    8.62  &   unres.  \\
J1330$-$3124a  &  13 30 05.10   &  $-$31 24 36.3   &    6.89  &   unres.  \\
J1330$-$3144a  &  13 30 05.20   &  $-$31 44 51.7   &    4.01  &   unres.  \\
J1330$-$3113   &  13 30 05.59   &  $-$31 13 44.8   &    1.79  &   unres.  \\
J1330$-$3215a  &  13 30 05.60   &  $-$32 15 54.3   &    1.81  &   unres.  \\
J1330$-$3143a  &  13 30 05.82   &  $-$31 43 42.2   &    4.02  &   unres.  \\
J1330$-$3102   &  13 30 07.17   &  $-$31 02 19.4   &   10.07  &    ext.   \\
J1330$-$3103   &  13 30 08.55   &  $-$31 03 46.2   &    1.08  &   unres.  \\
J1330$-$3144b  &  13 30 08.56   &  $-$31 44 02.8   &    3.26  &   unres.  \\
J1330$-$3116   &  13 30 09.79   &  $-$31 16 10.8   &    2.68  &   unres.  \\
J1330$-$3124b  &  13 30 10.06   &  $-$31 24 05.6   &    5.81  &   unres.  \\
J1330$-$3214   &  13 30 14.71   &  $-$32 14 46.3   &    1.94  &   unres.  \\
J1330$-$3122   &  13 30 19.06   &  $-$31 22 58.2   &  439.08  &   unres.  \\
J1330$-$3159   &  13 30 19.41   &  $-$31 59 40.0   &    2.05  &   unres.  \\
J1330$-$3201   &  13 30 21.69   &  $-$32 01 33.3   &    3.27  &   unres.  \\
J1330$-$3129   &  13 30 30.79   &  $-$31 29 59.3   &   14.13  &   unres.  \\
J1330$-$3153   &  13 30 33.03   &  $-$31 53 29.1   &    1.14  &   unres.  \\
J1330$-$3152   &  13 30 39.44   &  $-$31 52 11.0   &    1.56  &   unres.  \\
J1330$-$3105   &  13 30 40.88   &  $-$31 05 47.3   &    3.04  &   unres.  \\
J1330$-$3215b  &  13 30 41.11   &  $-$32 15 21.4   &    0.65  &   unres.  \\
J1330$-$3134   &  13 30 41.21   &  $-$31 34 17.4   &    3.86  &    ext.   \\
J1330$-$3141   &  13 30 42.14   &  $-$31 41 35.7   &    4.18  &   unres.  \\
J1330$-$3120   &  13 30 47.27   &  $-$31 20 30.4   &    2.22  &   unres.  \\
J1330$-$3143b  &  13 30 47.95   &  $-$31 43 40.2   &    7.77  &    ext.   \\
J1330$-$3148   &  13 30 48.03   &  $-$31 48 45.9   &    1.03  &   unres.  \\
J1330$-$3226   &  13 30 48.11   &  $-$32 26 07.5   &    2.61  &   unres.  \\
J1330$-$3145   &  13 30 48.27   &  $-$31 45 48.2   &    0.98  &   unres.  \\
J1330$-$3146   &  13 30 51.97   &  $-$31 46 00.0   &    1.98  &   unres.  \\
J1330$-$3209   &  13 30 52.00   &  $-$32 09 02.3   &    0.73  &   unres.  \\
J1330$-$3127   &  13 30 52.33   &  $-$31 27 01.1   &    1.04  &   unres.  \\
J1330$-$3204   &  13 30 55.62   &  $-$32 04 01.3   &    9.64  &   unres.  \\
J1330$-$3208   &  13 30 59.20   &  $-$32 08 54.2   &    0.51  &   unres.  \\
J1331$-$3144   &  13 31 00.60   &  $-$31 44 51.4   &    2.55  &   unres.  \\
J1331$-$3140   &  13 31 01.87   &  $-$31 40 35.3   &    0.60  &   unres.  \\
J1331$-$3127   &  13 31 05.38   &  $-$31 27 00.1   &    4.78  &   unres.  \\
J1331$-$3154a  &  13 31 06.71   &  $-$31 54 59.9   &    8.48  &    ext.   \\
J1331$-$3119   &  13 31 10.88   &  $-$31 19 27.6   &   14.52  &   unres.  \\
J1331$-$3139   &  13 31 12.00   &  $-$31 39 27.1   &    9.48  &   unres.  \\
J1331$-$3155a  &  13 31 14.14   &  $-$31 55 49.2   &    0.78  &  unres.   \\
\noalign{\smallskip}							  
\hline
\end{tabular}
\end{center}
\end{table*}
%
%
\setcounter{table}{2}
\begin{table*}
\caption[]{ Continued}
\begin{center}
\begin{tabular}{lcccc}
\hline\noalign{\smallskip}
Name  & RA$_{J2000}$   &  DEC$_{J2000}$ & S$_{1.4GHz}$   & Radio Morphology \\
VLA$-$&                &                &    (mJy)       &                  \\ 
\noalign{\smallskip}
\hline\noalign{\smallskip}
J1331$-$3124   &  13 31 14.63   &  $-$31 24 17.9   &    2.29   &  unres.   \\
J1331$-$3101   &  13 31 15.08   &  $-$31 01 38.6   &   12.72  &    ext.   \\
J1331$-$3135   &  13 31 16.85   &  $-$31 35 34.2   &    0.93   &  unres.   \\
J1331$-$3128   &  13 31 16.91   &  $-$31 28 20.1   &   22.28   &  unres.   \\
J1331$-$3113   &  13 31 18.79   &  $-$31 13 56.2   &    2.38   &  unres.   \\
J1331$-$3143   &  13 31 19.88   &  $-$31 43 50.2   &   14.53   &   ext.    \\
J1331$-$3108   &  13 31 25.72   &  $-$31 08 21.5   &    6.04   &  unres.   \\
J1331$-$3149a  &  13 31 27.54   &  $-$31 49 14.7   &    1.32   &  unres.   \\
J1331$-$3121a  &  13 31 29.66   &  $-$31 21 45.4   &   12.93   &  unres.   \\
J1331$-$3209a  &  13 31 31.57   &  $-$32 09 27.4   &    1.06   &   ext.    \\
J1331$-$3133a  &  13 31 37.66   &  $-$31 33 08.4   &    3.69   &  unres.   \\
J1331$-$3121b  &  13 31 39.06   &  $-$31 21 59.7   &    4.85   &   ext.    \\
J1331$-$3154b  &  13 31 39.29   &  $-$31 54 15.6   &    2.04   &   ext.    \\
J1331$-$3209b  &  13 31 41.58   &  $-$32 09 43.8   &    0.68   &  unres.   \\
J1331$-$3147a  &  13 31 41.96   &  $-$31 47 17.7   &    0.47   &  unres.   \\
J1331$-$3206   &  13 31 42.95   &  $-$32 06 38.2   &  129.85   &  unres.   \\
J1331$-$3147b  &  13 31 42.98   &  $-$31 47 45.8   &    0.47   &  unres.   \\
J1331$-$3221   &  13 31 43.37   &  $-$32 21 36.2   &   16.39   &   ext.    \\
J1331$-$3132   &  13 31 44.04   &  $-$31 32 52.5   &    1.34   &  unres.   \\
J1331$-$3116a  &  13 31 48.62   &  $-$31 16 23.8   &   51.46   &  unres.   \\
J1331$-$3133b  &  13 31 48.63   &  $-$31 33 07.8   &    1.04   &  unres.   \\
J1331$-$3148   &  13 31 50.22   &  $-$31 48 47.4   &    1.32   &  unres.   \\
J1331$-$3142   &  13 31 50.89   &  $-$31 42 49.7   &    2.13   &  unres.   \\
J1331$-$3155b  &  13 31 53.27   &  $-$31 55 33.4   &    2.06   &  unres.   \\
J1331$-$3116b  &  13 31 54.18   &  $-$31 16 44.6   &    8.74   &  unres.   \\
J1331$-$3154c  &  13 31 54.83   &  $-$31 54 01.2   &    1.19   &  unres.   \\
J1331$-$3149b  &  13 31 59.47   &  $-$31 49 19.8   &    1.82   &   ext.    \\
J1332$-$3136a  &  13 32 00.29   &  $-$31 36 26.4   &    0.40   &  unres.   \\
J1332$-$3112   &  13 32 02.35   &  $-$31 12 44.8   &    4.53   &  unres.   \\
J1332$-$3146a  &  13 32 02.78   &  $-$31 46 50.0   &   15.11   &   ext.    \\
J1332$-$3152a  &  13 32 05.13   &  $-$31 52 30.4   &    0.49   &  unres.   \\
J1332$-$3141a  &  13 32 05.34   &  $-$31 41 23.4   &   11.32   &  unres.   \\
J1332$-$3228   &  13 32 11.04   &  $-$32 28 23.7   &   11.17   &  unres.   \\
J1332$-$3123a  &  13 32 14.42   &  $-$31 23 59.8   &    3.02   &  unres.   \\
J1332$-$3131a  &  13 32 15.98   &  $-$31 31 25.7   &    0.83   &  unres.   \\
J1332$-$3152b  &  13 32 17.52   &  $-$31 52 49.6   &   15.78   &  unres.   \\
J1332$-$3123b  &  13 32 27.65   &  $-$31 23 49.5   &   21.09   &  unres.   \\
J1332$-$3151   &  13 32 29.84   &  $-$31 51 01.8   &    0.86   &  unres.   \\
J1332$-$3141b  &  13 32 31.71   &  $-$31 41 54.5   &    3.24   &  unres.   \\
J1332$-$3131b  &  13 32 35.38   &  $-$31 31 41.8   &    1.79   &  unres.   \\
J1332$-$3135   &  13 32 37.56   &  $-$31 35 55.5   &    0.76   &  unres.   \\
J1332$-$3201   &  13 32 40.51   &  $-$32 01 56.9   &   32.19   &   ext.    \\
J1332$-$3124   &  13 32 42.47   &  $-$31 24 52.4   &    0.96   &  unres.   \\
J1332$-$3155   &  13 32 43.25   &  $-$31 55 11.0   &    0.58   &  unres.   \\
J1332$-$3146b  &  13 32 43.92   &  $-$31 46 53.2   &    1.41   &  unres.   \\
J1332$-$3142a  &  13 32 44.59   &  $-$31 42 44.8   &    0.34   &  unres.   \\
J1332$-$3158   &  13 32 44.98   &  $-$31 58 24.7   &   17.64   &  unres.   \\
J1332$-$3136b  &  13 32 46.59   &  $-$31 36 58.5   &    1.92   &  unres.   \\
J1332$-$3156   &  13 32 50.23   &  $-$31 56 11.5   &    0.76   &  unres.   \\
J1332$-$3142b  &  13 32 51.48   &  $-$31 42 37.3   &    0.40   &  unres.   \\
J1332$-$3144   &  13 32 55.07   &  $-$31 44 14.4   &    0.90   &  unres.   \\
J1332$-$3128   &  13 32 55.69   &  $-$31 28 12.3   &    1.12   &  unres.   \\
J1332$-$3134   &  13 32 56.48   &  $-$31 34 52.5   &    0.72   &  unres.   \\ 
J1332$-$3146c  &  13 32 57.03   &  $-$31 46 09.4   &    0.47   &  unres.   \\ 
J1332$-$3123c  &  13 32 57.21   &  $-$31 23 53.2   &    6.56   &  unres.   \\ 
J1332$-$3148   &  13 32 57.58   &  $-$31 48 04.2   &    2.41   &  unres.   \\ 
\noalign{\smallskip}
\hline
\end{tabular}
\end{center}
\end{table*}
%
%
\setcounter{table}{2}
\begin{table*}
\caption[]{ Continued}
\begin{center}
\begin{tabular}{lcccc}
\hline\noalign{\smallskip}
Name  & RA$_{J2000}$   &  DEC$_{J2000}$ & S$_{1.4GHz}$   & Radio Morphology \\
VLA$-$&                &                &    (mJy)       &                  \\ 
\noalign{\smallskip}
\hline\noalign{\smallskip}
J1332$-$3125   &  13 32 59.08   &  $-$31 25 17.5   &    1.01   &  unres.  \\
J1333$-$3119a  &  13 33 00.78   &  $-$31 19 19.8   &    3.09   &   ext.   \\
J1333$-$3144   &  13 33 01.47   &  $-$31 44 12.9   &    1.49   &  unres.  \\
J1333$-$3139a  &  13 33 04.27   &  $-$31 39 04.0   &    1.50   &  unres.  \\
J1333$-$3123   &  13 33 05.05   &  $-$31 23 55.5   &    2.20   &  unres.  \\
J1333$-$3145a  &  13 33 07.07   &  $-$31 45 47.0   &    1.07   &  unres.  \\
J1333$-$3147a  &  13 33 08.97   &  $-$31 47 33.5   &    0.80   &  unres.  \\
J1333$-$3146   &  13 33 11.80   &  $-$31 46 50.6   &    0.52   &  unres.  \\
J1333$-$3143a  &  13 33 15.81   &  $-$31 43 08.7   &    0.52   &  unres.  \\
J1333$-$3153   &  13 33 17.71   &  $-$31 53 24.6   &    1.54   &  unres.  \\
J1333$-$3139b  &  13 33 17.83   &  $-$31 39 09.2   &    1.56   &  unres.  \\
J1333$-$3134   &  13 33 22.64   &  $-$31 34 42.0   &    0.62   &  unres.  \\
J1333$-$3124   &  13 33 23.02   &  $-$31 24 10.0   &    0.52   &  unres.  \\
J1333$-$3143b  &  13 33 24.58   &  $-$31 43 06.7   &    0.72   &  unres.  \\
J1333$-$3125   &  13 33 27.32   &  $-$31 25 16.7   &    0.90   &  unres.  \\
J1333$-$3156   &  13 33 30.11   &  $-$31 56 30.9   &    0.89   &  unres.  \\
J1333$-$3141   &  13 33 32.00   &  $-$31 41 18.1   &  109.33   & ext.(HT) \\
J1333$-$3138   &  13 33 32.11   &  $-$31 38 20.7   &    3.76   &  unres.  \\
J1333$-$3130   &  13 33 37.29   &  $-$31 30 46.5   &   43.10   &  unres.  \\
J1333$-$3147b  &  13 33 39.36   &  $-$31 47 06.4   &    0.33   &  unres.  \\
J1333$-$3145b  &  13 33 40.10   &  $-$31 45 22.4   &    0.69   &  unres.  \\
J1333$-$3119b  &  13 33 40.63   &  $-$31 19 53.5   &    5.94   &  unres.  \\
J1333$-$3154   &  13 33 40.90   &  $-$31 54 15.3   &    0.45   &  unres.  \\
J1333$-$3142   &  13 33 41.35   &  $-$31 42 29.0   &    0.50   &  unres.  \\
J1333$-$3135   &  13 33 41.39   &  $-$31 35 50.0   &    3.14   &   ext.   \\
J1333$-$3139c  &  13 33 46.24   &  $-$31 39 27.3   &    1.76   &  unres.  \\
J1333$-$3128   &  13 33 49.64   &  $-$31 28 31.2   &    3.23   &   ext.   \\
J1333$-$3139d  &  13 33 51.28   &  $-$31 39 36.5   &    1.02   &  unres.  \\
J1333$-$3119c  &  13 33 52.96   &  $-$31 19 56.9   &   13.94   &  unres.  \\
J1333$-$3129   &  13 33 57.94   &  $-$31 29 05.8   &    1.09   &  unres.  \\
J1333$-$3158   &  13 33 58.72   &  $-$31 58 11.9   &    1.00   &  unres.  \\
J1333$-$3132   &  13 33 59.12   &  $-$31 32 52.0   &    0.60   &  unres.  \\
J1334$-$3146   &  13 34 03.59   &  $-$31 46 31.5   &    0.37   &  unres.  \\
J1334$-$3119a  &  13 34 04.08   &  $-$31 19 25.5   &    3.78   &  unres.  \\
J1334$-$3143   &  13 34 06.49   &  $-$31 43 37.4   &    0.97   &  unres.  \\
J1334$-$3131a  &  13 34 07.80   &  $-$31 31 02.9   &    1.12   &  unres.  \\
J1334$-$3128   &  13 34 08.32   &  $-$31 28 36.2   &   33.63   &  unres.  \\
J1334$-$3136a  &  13 34 10.54   &  $-$31 36 57.0   &    3.45   &   ext.   \\
J1334$-$3153   &  13 34 11.15   &  $-$31 53 36.0   &    0.66   &  unres.  \\
J1334$-$3123   &  13 34 12.49   &  $-$31 23 54.3   &    6.79   &  unres.  \\
J1334$-$3126a  &  13 34 13.43   &  $-$31 26 39.1   &    4.91   &  unres.  \\
J1334$-$3149a  &  13 34 13.84   &  $-$31 49 48.6   &    5.51   &  unres.  \\
J1334$-$3142   &  13 34 15.56   &  $-$31 42 24.7   &    0.40   &  unres.  \\
J1334$-$3149b  &  13 34 17.71   &  $-$31 49 07.8   &    1.21   &  unres.  \\
J1334$-$3131b  &  13 34 19.03   &  $-$31 31 08.8   &    1.28   &  unres.  \\
J1334$-$3119b  &  13 34 22.11   &  $-$31 19 08.4   &   41.70   &  unres.  \\
J1334$-$3139a  &  13 34 22.47   &  $-$31 39 06.8   &   16.77   &  unres.  \\
J1334$-$3137   &  13 34 22.82   &  $-$31 37 08.3   &    0.74   &  unres.  \\
J1334$-$3136b  &  13 34 29.22   &  $-$31 36 44.7   &    1.60   &  unres.  \\
J1334$-$3141a  &  13 34 36.65   &  $-$31 41 01.6   &    2.11   &  unres.  \\
J1334$-$3132   &  13 34 37.36   &  $-$31 32 47.9   &   26.48   &   ext.   \\
J1334$-$3124   &  13 34 40.05   &  $-$31 24 56.5   &    5.25   &  unres.  \\
J1334$-$3126b  &  13 34 40.57   &  $-$31 26 32.4   &    2.71   &  unres.  \\
J1334$-$3141b  &  13 34 46.19   &  $-$31 41 35.7   &    0.95   &  unres.  \\
J1334$-$3155   &  13 34 46.64   &  $-$31 55 20.2   &    1.38   &  unres.  \\
\noalign{\smallskip}							  
\hline
\end{tabular}
\end{center}
\end{table*}
%
%
\setcounter{table}{2}
\begin{table*}
\caption[]{ Continued}
\begin{center}
\begin{tabular}{lcccc}
\hline\noalign{\smallskip}
Name  & RA$_{J2000}$   &  DEC$_{J2000}$ & S$_{1.4GHz}$   & Radio Morphology \\
VLA$-$&                &                &    (mJy)       &                  \\ 
\noalign{\smallskip}
\hline\noalign{\smallskip}
J1334$-$3125   &  13 34 51.01   &  $-$31 25 25.5   &    3.19   &  unres.   \\
J1334$-$3151   &  13 34 52.21   &  $-$31 51 10.1   &    3.84   &  unres.   \\
J1334$-$3139b  &  13 34 54.11   &  $-$31 39 14.9   &    0.77   &  unres.   \\
J1335$-$3139   &  13 35 02.97   &  $-$31 39 11.6   &   16.71   &  unres.   \\
J1335$-$3130   &  13 35 08.78   &  $-$31 30 31.3   &    1.63   &  unres.   \\
J1335$-$3133   &  13 35 12.17   &  $-$31 33 37.3   &    2.21   &  unres.   \\
J1335$-$3134   &  13 35 18.98   &  $-$31 34 41.9   &    2.77   &  unres.   \\
\noalign{\smallskip}
\hline
\end{tabular}
\end{center}
\end{table*}
%
%

As clear from Table 3
most of the radio sources detected at 1.4 GHz  are unresolved; only 
23 radio sources are extended, i.e. 13\% of the total. 

Our observations confirm the presence of a radio halo at the centre of  
A3562 (not listed in Table 3),
encompassing the head--tail radio galaxy J1333$-$3141. 
The halo has a flux density of $S_{1.4GHz}$ = 20 mJy, a 
low surface brightness ($\le$ 0.175 mJy b$^{-1}$) and an irregular 
shape, 
with largest angular size of $\sim$ 8$^{\prime}$.
The properties and the origin of this source are discussed in a different
paper (V2003).

We note that the region presented in this paper (A3558--C)
partly overlaps with the area covered in the ATCA survey presented in
V2000. We carefully checked all sources in the common region, 
and found that all differences can be accounted for either 
by the different u--v coverage, resolution and/or sensitivity of the 
two sets of observations. Due to the different resolutions 
($\sim 10^{\prime\prime}\times 5^{\prime\prime}$ for the ATCA observations), 
in a number of cases the radio emission detected with the present 
observations is actually a blend of two or more sources in the ATCA 22 cm 
sample.Those cases of source blending relevant to the discussion
of this paper are briefly commented in the Appendix.

\subsection{Radio source counts}

We computed the source counts for our radio sample in 
A3558--C, in order to test if the optical galaxy overdensity 
and the ongoing merging scenario proposed here reflect
into a higher number of radio sources with respect to the 
field radio sources counts.

Since the sensitivity in the final images of the
four fields covered by our observations is not uniform due
to the VLA primary beam attenuation, the 1.4 GHz sample 
is not complete to the flux density limit of 0.25 mJy. 
For this analysis we therefore considered only the 
radio sources with S$_{1.4GHz} \ge 0.50$ mJy within
a radius of 15.5 arcmin from the centre of each field.
At such distance the primary beam attenuation of the VLA at 1.4 GHz
is reduced by a factor of two, and sources with flux density 
S$_{1.4GHz} \ge 0.50$ mJy are seen as sources with S$_{1.4GHz}\ge 0.25$ 
mJy before the correction. At the distance of A3558--C this
limit implies a radio power logP$_{1.4GHz}$(W Hz$^{-1}$)=
21.13. 

Our results are in very good agreement with the radio source 
field counts (Prandoni et al. \cite{prandoni01}) and 
with the statistical analysis over the whole A3558 cluster chain 
carried out in V2000.
This implies that even in an overdense and merging region 
such as the one considered here, the radio source 
counts are dominated by field sources, at least for 
logP$_{1.4GHz}$ (W Hz$^{-1}$) \gtsim  21.

\section{Optical Identifications}\label{sec:ident}

In order to find as many optical counterparts as possible,
we cross--correlated the radio positions of all sources in
our sample with four different optical catalogues.
Our reference catalogue is the COSMOS/UKST Southern Sky Object 
Catalogue (Yentis et al. \cite{yentis92}), limited to $b_J$ = 19.5;
then for fainter magnitudes we considered also the 
SuperCOSMOS/USKT Southern Sky Object Catalogue (Hambly et al. 
\cite{hambly01}), the APM Catalogue (Maddox et al. 
\cite{maddox90}) and finally the MGP catalogue (Metcalfe et al. 
\cite{metcalfe94}). 

All these catalogues have a claimed positional accuracy of 
$\sim$ 0.25$^{\prime \prime}$, but given the errors which could 
be introduced by transforming the sky image on the plate frame, 
we adopted a mean optical positional uncertainty of 1.5 arcsec 
(Unewisse et al. \cite{unewisse93}).

The radio positional error depends on the beam size and on the 
source flux density (Prandoni et al. 2000).
With the parameters of our observations we estimate an average
position uncertainty of 2.5 arcsec both in right ascension and 
declination. For the faintest sources in the sample 
(5$\sigma$) the positional error raises to 
$\sim 4.2^{\prime\prime} \times 3.5^{\prime\prime}$ along the
beam axis (HPBW/(2$\times$SNR), being SNR the signal--to--noise
ratio).

In order to make sure that no identification was missed, we
examined also the photometric catalogue of optical galaxies in this 
region in Metcalfe et al. (\cite{metcalfe94}) and
we overplotted all the radio sources in our sample on the optical
red Digitized Sky Survey DSS--2 images and carried out a careful
visual inspection.

Given the uncertainty in the radio and optical positions, 
in order to estimate the reliability of the optical identifications,
we adopted the parameter ${\cal R}$, defined as:

$${\cal R}^2 = \frac{\Delta_{r-o}^2}{\sigma^2_o + \sigma^2_r}$$

\noindent 
where $\Delta_{r-o}$ is the positional offset between
radio and optical coordinates and $\sigma_o$= 1.5 arcsec and 
$\sigma_r$= 2.5 arcsec are the optical
and radio position errors respectively.

We considered reliable identifications all
matches with ${\cal R} \le$ 3. Given the extent of the radio emission we 
found ${\cal R} >$ 3 for a number of sources, but we considered them 
reliable identifications since the optical counterpart falls within the 
radio isophotes (further details are given in Appendix).

The reliability ({\it rel}) and completeness ({\it comp}) of our sample of 
identified radio sources was tested following the method suggested by de 
Ruiter et
al. (1977). We note that the reliability represents the 
fraction of true, i.e. non spurious, optical identifications in the sample, 
while the completeness provides  the fraction of true identifications found 
with respect to the total number of IDs in the sample.
For $\cal R$ = 3 we found {\it rel} = 96.3\% and {\it comp} = 97.8\%; while
for $\cal R$ = 2 we have {\it rel} = 99\% and {\it comp} = 68.3\%. 
We note that completeness drops considerably going from $\cal R$ = 3 to
$\cal R$ = 2. For this reason, and in the light of the puropse of the 
present paper (see Sections 6.1, 6.3, 
7.1 and 7.2), we consider $\cal R$ = 3 as the most appropriate choice.

The list of the radio-optical identification is reported in Table 4, 
where we give:

\parn $-$ column 1 : radio and optical name, where \# stands for
optical counterparts from Bardelli et al. (\cite{bardelli94}, \cite{bardelli98b}) and 
MGP94 from Metcalfe et al. (\cite{metcalfe94}); 

\parn $-$ columns 2 and 3: J2000 radio and optical coordinates;

\parn $-$ column 4: radio flux density at 1.4 GHz and $b_J$ magnitude; 

\parn $-$ column 5: radio morphology and optical type, where E = elliptical, 
S = spiral, IS = interacting system, qso = quasar candidate;

\parn $-$ column 6: radio power at 1.4 GHz and radial velocity;

\parn $-$ column 7: ${\cal R}$ parameter and (B--R) colours for the 
optical counterparts in the Shapley Concentration taken from Metcalfe et al. 
(\cite{metcalfe94}).

%
%
\setcounter{table}{3}
\begin{table*}
\label{tab:optic}
\caption[]{Optical Identifications}
\begin{center}
\begin{tabular}{lllrccc}
\hline\noalign{\smallskip}
Radio Name    & RA$_{J2000}$  & DEC$_{J2000}$ & $S_{1.4GHz}$ & Radio type & logP$_{1.4GHz}$ &  ${\cal R}$ \\
VLA$-$        &               &               &  (mJy)       &            & h$^{-2}$ (W Hz$^{-1}$)  &             \\
Opt. Name     & RA$_{J2000}$  & DEC$_{J2000}$ &   $b_J$      & Opt.type   &      v          &    B--R     \\
              &               &               &              &            & (km s$^{-1}$)   &             \\  
\\
\noalign{\smallskip}
\hline\noalign{\smallskip}
J1329$-$3126  &  13 29 00.02  & $-$31 26 26.0   &   81.92    &   ext.    &    $-$     &   2.88   \\
              &  13 29 00.22  & $-$31 26 34.0   &   18.96    &   qso     &    $-$     &   $-$    \\
              &               &                 &            &           &            &          \\
J1329$-$3119a &  13 29 49.32  & $-$31 19 37.3   &    6.02    &  unres.   &    $-$     &   1.58   \\
              &  13 29 49.02  & $-$31 19 39.9   &   15.09    &   $-$     &    $-$     &   $-$    \\
              &               &                 &            &           &            &          \\ 
J1329$-$3122a &  13 29 50.76  & $-$31 22 55.6   &   22.43    &  unres.   &   24.00    &   1.17 $\diamond$\\
$\#$10178     &  13 29 50.8   & $-$31 22 59     &   19.08    &    E      &   58755    &   $-$    \\
              &               &                 &            &           &            &          \\
J1329$-$3057  &  13 29 51.22  & $-$30 57 47.2   &   65.38    &  unres.   &    $-$     &   1.34   \\
              &  13 29 51.44  & $-$30 57 49.8   &   22.15    &   $-$     &    $-$     &   $-$    \\
              &               &                 &            &           &            &          \\
J1329$-$3119b &  13 29 55.02  & $-$31 19 57.7   &    4.53    &  unres.   &   22.07    &   0.10   \\
$\#$10174     &  13 29 55.00  & $-$31 19 57.6   &   15.15    &     S     &   14799    &   2.05   \\
              &               &                 &            &           &            &          \\
J1330$-$3124a &  13 30 05.10  & $-$31 24 36.3   &    6.89    &  unres.   &   22.20    &   0.41   \\
$\#$10318     &  13 30 05.05  & $-$31 24 37.2   &   16.05    &     S     &   13770    &   1.86   \\
              &               &                 &            &           &            &          \\
J1330$-$3113  & 13 30 05.59   & $-$31 13 44.8   &    1.79    &  unres.   &   21.76    &   3.70   \\
$\#10313$     & 13 30 05.09   & $-$31 13 53.5   &   17.44    &     S     &   15942    &   1.53   \\
              &               &                 &            &           &            &          \\
J1330$-$3144b & 13 30 08.56   & $-$31 44 02.8   &    3.26    &  unres.   &   23.06    &   2.50   \\
 $\#$10332    & 13 30 09.05   & $-$31 43 59.0   &   19.34    &   $-$     &   52606    &   $-$    \\
              &               &                 &            &           &            &          \\
J1330$-$3124b & 13 30 10.06   & $-$31 24 05.6   &    5.81    &  unres.   &    $-$     &   1.58   \\
              & 13 30 10.16   & $-$31 24 01.2   &   22.13    &   $-$     &    $-$     &   $-$    \\
              &               &                 &            &           &            &          \\
J1330$-$3122  & 13 30 19.06   & $-$31 22 58.2   &  439.08    &  unres.   &    $-$     &   0.31   \\
              & 13 30 19.1    & $-$31 22 59     &   18.46    &    qso    &    $-$     &   $-$    \\ 
              &               &                 &            &           &            &          \\
J1330$-$3153  & 13 30 33.03   & $-$31 53 29.1   &    1.14    &  unres.   &    $-$     &   1.13   \\
              & 13 30 32.85   & $-$31 53 31.5   &   18.03    &   $-$     &    $-$     &   $-$    \\
              &               &                 &            &           &            &          \\
J1330$-$3134  & 13 30 41.21   & $-$31 34 17.4   &    3.86    &    ext.   &   22.06    &   0.41   \\
$\#$10725     & 13 30 41.29   & $-$31 34 16.9   &   16.40    &     S     &   15524    &   1.72   \\
              &               &                 &            &           &            &          \\
J1330$-$3143b & 13 30 47.95   & $-$31 43 40.2   &    7.77    &    ext.   &   22.25    &   2.30 $\diamond$\\
$\#$10869     & 13 30 47.95   & $-$31 43 33.5   &   15.58    &     S     &   13894    &   $-$    \\
              &               &                 &            &           &            &          \\
J1330$-$3146  & 13 30 51.97   & $-$31 46 00.0   &    1.98    &  unres.   &   21.66    &   2.74   \\
$\#$10871     & 13 30 52.13   & $-$31 46 07.8   &   17.79    &     E     &   13832    &   1.86   \\
              &               &                 &            &           &            &          \\
J1330$-$3209  & 13 30 52.00   & $-$32 09 02.3   &    0.73    &  unres.   &   21.17    &   2.09   \\
$\#$10888     & 13 30 52.08   & $-$32 08 56.3   &   16.55    &     S     &   12932    &   1.78   \\
              &               &                 &            &           &            &          \\
J1330$-$3127  & 13 30 52.33   & $-$31 27 01.1   &    1.04    &  unres.   &   21.44    &   4.70 $\diamond$\\
$\#$10986     & 13 30 52.59   & $-$31 27 14.1   &   17.76    &     E     &   14618    &   1.94   \\
              &               &                 &            &           &            &          \\
J1331$-$3144  & 13 31 00.60   & $-$31 44 51.4   &    2.55    &  unres.   &   21.67    &   0.75   \\
$\#$11000     & 13 31 00.65   & $-$31 44 53.6   &   16.98    &     S     &   12141    &   1.68   \\
              &               &                 &            &           &            &          \\
J1331$-$3140  & 13 31 01.87   & $-$31 40 35.3   &    0.60    &  unres.   &   21.10    &   0.55   \\
$ \#$10996    & 13 31 01.91   & $-$31 40 36.8   &   18.22    &     E     &   13178    &   1.67   \\
              &               &                 &            &           &            &          \\
J1331$-$3139  & 13 31 12.00   & $-$31 39 27.1   &    9.48    &  unres.   &    $-$     &   3.81 $\diamond$\\
              & 13 31 11.6    & $-$31 39 37     &   17.97    &   qso     &    $-$     &   $-$    \\  
              &               &                 &            &           &            &          \\
\noalign{\smallskip}
\hline
\end{tabular}
\end{center}
\end{table*}
%
%
\setcounter{table}{3}
\begin{table*}
\caption[]{Optical Identifications. Continued}
\begin{center}
\begin{tabular}{lllrccc}
\hline\noalign{\smallskip}
Radio Name    & RA$_{J2000}$  & DEC$_{J2000}$ & $S_{1.4GHz}$ & Radio type &logP$_{1.4GHz}$ &  ${\cal R}$ \\
VLA$-$        &               &               &    (mJy)     &            & h$^{-2}$ (W Hz$^{-1}$) &             \\
Opt. Name     & RA$_{J2000}$  & DEC$_{J2000}$ &   $b_J$      & Opt.type   &      v         &    B--R     \\
              &               &               &              &            & (km s$^{-1}$)  &             \\  
\\
\noalign{\smallskip}
\hline\noalign{\smallskip}
J1331$-$3155a & 13 31 14.14   & $-$31 55 49.2   &    0.78    &  unres.   &    $-$     &   1.34   \\
              & 13 31 14.38   & $-$31 55 46.8   &   20.47    &   $-$     &    $-$     &   $-$    \\
              &               &                 &            &           &            &          \\
J1331$-$3135  & 13 31 16.85   & $-$31 35 34.2   &    0.93    &  unres.   &   21.39    &   2.47   \\
 MGP94 3226   & 13 31 17.32   & $-$31 35 38.1   &   17.13    &     E     &   14578    &   1.53   \\
              &               &                 &            &           &            &          \\ 
J1331$-$3128  & 13 31 16.91   & $-$31 28 20.1   &   22.28    &  unres.   &    $-$     &   2.98   \\
              & 13 31 17.3    & $-$31 28 13     &   17.72    &   qso     &    $-$     &   $-$    \\
              &               &                 &            &           &            &          \\
J1331$-$3113  & 13 31 18.79   & $-$31 13 56.2   &    2.38    &  unres.   &   21.83    &   4.63   \\
$\#$11243     & 13 31 19.38   & $-$31 13 45.0   &   16.35    &     S     &   15710    &   1.97   \\ 
              &               &                 &            &           &            &          \\
J1331$-$3143  & 13 31 19.88   & $-$31 43 50.2   &   14.53    &   ext.    &    $-$     &   2.71 $\diamond$\\
              & 13 31 19.84   & $-$31 43 42.3   &   19.36    &    $-$    &    $-$     &   $-$    \\
              &               &                 &            &           &            &          \\ 
J1331$-$3149a & 13 31 27.54   & $-$31 49 14.7   &    1.32    &  unres.   &   21.44    &   0.62   \\
MGP94 3303    & 13 31 27.4    & $-$31 49 14.5   &   15.64    &     E     &   12928    &   2.05   \\
              &               &                 &            &           &            &          \\
J1331$-$3133a & 13 31 37.66   & $-$31 33 08.4   &    3.69    &  unres.   &   21.95    &   0.45   \\
$\#$11515     & 13 31 37.57   & $-$31 33 07.7   &   15.71    &     E     &   14205    &   $-$    \\
              &               &                 &            &           &            &          \\
J1331$-$3206  & 13 31 42.95   & $-$32 06 38.2   &  129.85    &  unres.   &    $-$     &   2.54   \\
              & 13 31 42.8    & $-$32 06 31     &   17.20    &   qso     &    $-$     &   $-$    \\
              &               &                 &            &           &            &          \\
J1331$-$3148  & 13 31 50.22   & $-$31 48 47.4   &    1.32    &  unres.   &   21.46    &   3.09   \\
$\#$11639     & 13 31 50.77   & $-$31 48 41.7   &   16.45    &     E     &   13343    &   $-$    \\
              &               &                 &            &           &            &          \\
J1331$-$3142  & 13 31 50.89   & $-$31 42 49.7   &    2.13    &  unres.   &    $-$     &   0.82   \\
              & 13 31 51.01   & $-$31 42 47.9   &   22.80    &   $-$     &    $-$     &   $-$    \\
              &               &                 &            &           &            &          \\
J1331$-$3155b & 13 31 53.27   & $-$31 55 33.4   &    2.06    &  unres.   &    $-$     &   0.99 $\diamond$\\
$\#$11647     & 13 31 53.22   & $-$31 55 36.2   &   19.08    &     E     &    $-$     &   $-$    \\
              &               &                 &            &           &            &          \\
J1331$-$3154c & 13 31 54.83   & $-$31 54 01.2   &    1.19    &  unres.   &   21.40    &   1.13   \\
$\#$11644     & 13 31 54.77   & $-$31 54 04.4   &   17.52    &     E     &   13028    &   1.84   \\
              &               &                 &            &           &            &          \\ 
J1331$-$3149b & 13 31 59.47   & $-$31 49 19.8   &    1.82    &    ext.   &   21.58    &   1.54   \\
$ \#$11751    & 13 31 59.72   & $-$31 49 23.1   &   17.46    &     S     &   12963    &   $-$    \\
              &               &                 &            &           &            &          \\
J1332$-$3136a & 13 32 00.29   & $-$31 36 26.4   &    0.40    &  unres.   &    $-$     &   1.65   \\
              & 13 32 00.65   & $-$31 36 24.8   &   19.19    &    $-$    &    $-$     &   $-$    \\
              &               &                 &            &           &            &          \\
J1332$-$3146a & 13 32 02.78   & $-$31 46 50.0   &   15.11    &    ext.   &   22.52    &   0.99   \\
 $\#$11744    & 13 32 03.02   & $-$31 46 49.9   &   14.96    &     E     &   13107    &   2.07   \\
              &               &                 &            &           &            &          \\
J1332$-$3152a & 13 32 05.13   & $-$31 52 30.4   &    0.49    &  unres.   &   21.04    &   1.99   \\
$\#$11753     & 13 32 05.58   & $-$31 52 30.0   &   16.41    &     S     &   13634    &   1.67   \\
              &               &                 &            &           &            &          \\
J1332$-$3141a & 13 32 05.34   & $-$31 41 23.4   &   11.32    &  unres.   &    $-$     &   0.72 $\diamond$ \\
              & 13 32 05.47   & $-$31 41 24.7   &   19.85    &   $-$     &    $-$     &   $-$    \\
              &               &                 &            &           &            &          \\
J1332$-$3131a & 13 32 15.98   & $-$31 31 25.7   &    0.83    &  unres.   &    $-$     &   2.40   \\
 $\#$11872    & 13 32 15.89   & $-$31 31 32.6   &   18.08    &     S     &    $-$     &   $-$    \\
              &               &                 &            &           &            &          \\
J1332$-$3152b & 13 32 17.52   & $-$31 52 49.6   &   15.78    &  unres.   &    $-$     &   0.75   \\
              & 13 32 17.68   & $-$31 52 50.4   &   22.46    &   $-$     &    $-$     &   $-$    \\   
              &               &                 &            &           &            &          \\
\noalign{\smallskip}
\hline
\end{tabular}
\end{center}
\end{table*}
%
%
\setcounter{table}{3}
\begin{table*}
\caption[]{Optical Identifications. Continued}
\begin{center}
\begin{tabular}{lllrccc}
\hline\noalign{\smallskip}
Radio Name    & RA$_{J2000}$  & DEC$_{J2000}$ & $S_{1.4GHz}$ & Radio type &logP$_{1.4GHz}$ &  ${\cal R}$ \\
VLA$-$        &               &               &    (mJy)     &            & h$^{-2}$ (W Hz$^{-1}$) &             \\
Opt. Name     & RA$_{J2000}$  & DEC$_{J2000}$ &    $b_J$     & Opt.type   &      v         &  B--R       \\
              &               &               &              &            & (km s$^{-1}$)  &             \\  
\\
\noalign{\smallskip}
\hline\noalign{\smallskip}
J1332$-$3131b  & 13 32 35.38   & $-$31 31 41.8   &    1.79    &  unres.   &   21.71   &   1.61   \\
$\#$12072     & 13 32 35.24   & $-$31 31 46.2   &   17.28    &     E     &   15193   &   1.36   \\
              &               &                 &            &           &           &          \\
J1332$-$3135  & 13 32 37.56   & $-$31 35 55.5   &    0.76    &  unres.   &    $-$    &   1.85   \\
              & 13 32 37.98   & $-$31 35 54.9   &   15.80    &   $-$     &    $-$    &   $-$    \\
              &               &                 &            &           &           &          \\
J1332$-$3155  & 13 32 43.25   & $-$31 55 11.0   &    0.58    &  unres.   &   21.18   &   2.19   \\
 $ \#$12200   & 13 32 42.85   & $-$31 55 14.8   &   16.96    &     E     &   14483   &   1.73   \\
              &               &                 &            &           &           &          \\
J1332$-$3146b & 13 32 43.92   & $-$31 46 53.2   &    1.41    &  unres.   &   21.46   &   2.92   \\
$\#$12195     & 13 32 43.87   & $-$31 47 01.7   &   17.33    &     E     &   12836   &   1.58   \\
              &               &                 &            &           &           &          \\
J1332$-$3142a & 13 32 44.59   & $-$31 42 44.8   &    0.34    &  unres.   &    $-$    &   0.75   \\
              & 13 32 44.75   & $-$31 42 57.0   &   19.13    &   $-$     &    $-$    &   $-$    \\
              &               &                 &            &           &           &          \\
J1332$-$3136b & 13 32 46.59   & $-$31 36 58.5   &    1.92    &  unres.   &   21.56   &   1.82   \\
 $\#$12190    & 13 32 46.44   & $-$31 37 03.4   &   16.88    &     E     &   12317   &   1.34   \\
              &               &                 &            &           &           &          \\
J1332$-$3156  & 13 32 50.23   & $-$31 56 11.5   &    0.76    &  unres.   &    $-$    &   0.89   \\
              & 13 32 50.33   & $-$31 56 09.2   &   20.92    &   $-$     &    $-$    &   $-$    \\ 
              &               &                 &            &           &           &          \\
J1332$-$3134  & 13 32 56.48   & $-$31 34 52.5   &    0.72    &  unres.   &    $-$    &   0.82   \\
              & 13 32 56.34   & $-$31 34 51.0   &   19.63    &   $-$     &    $-$    &   $-$    \\
              &               &                 &            &           &           &          \\
J1333$-$3144  & 13 33 01.47   & $-$31 44 12.9   &    1.49    &  unres.   &    $-$    &   2.50   \\
$\#$12286     & 13 33 01.24   & $-$31 44 19.6   &   18.50    &     S     &    $-$    &   $-$    \\
              &               &                 &            &           &           &          \\
J1333$-$3139a & 13 33 04.27   & $-$31 39 04.0   &    1.50    &  unres.   &   21.57   &   0.82   \\
  $\#$12372   & 13 33 04.08   & $-$31 39 04.6   &   18.32    &     S     &   14120   &   1.85   \\
              &               &                 &            &           &           &          \\
J1333$-$3123  & 13 33 05.05   & $-$31 23 55.5   &    2.20    &  unres.   &    $-$    &   0.99   \\
              & 13 33 05.12   & $-$31 23 58.2   &   20.24    &   $-$     &    $-$    &   $-$    \\
              &               &                 &            &           &           &          \\
J1333$-$3124  & 13 33 23.02   & $-$31 24 10.0   &    0.52    &  unres.   &   21.04   &   5.04   \\
$\#$12572     & 13 33 23.46   & $-$31 24 23.6   &   17.32    &   $-$     &   13103   &   1.94   \\        
              &               &                 &            &           &           &          \\
J1333$-$3141  & 13 33 32.00   & $-$31 41 18.1   &  109.33    &    HT     &   23.43   &   6.58   \\
 MGP94 4108    & 13 33 31.5    & $-$31 41 00     &   17.25    &     E     &   14438   &   2.09   \\
              &               &                 &            &           &           &          \\
J1333$-$3130  & 13 33 37.29   & $-$31 30 46.5   &   43.10    &  unres.   &    $-$    &   0.45   \\
              & 13 33 37.2    & $-$31 30 46     &   22.42    &   $-$     &    $-$    &   $-$    \\
              &               &                 &            &           &           &          \\
J1333$-$3119b & 13 33 40.63   & $-$31 19 53.5   &    5.94    &  unres.   &    $-$    &   2.23   \\
  $\#$12661   & 13 33 40.60   & $-$31 19 60.0   &   16.64    &     IS    &    $-$    &   $-$    \\
              &               &                 &            &           &           &          \\
J1333$-$3135  & 13 33 41.39   & $-$31 35 50.0   &    3.14    &   ext.    &    $-$    &   0.31   \\
              & 13 33 41.36   & $-$31 35 49.2   &   21.96    &    $-$    &    $-$    &   $-$    \\
              &               &                 &            &           &           &          \\
J1333$-$3128  & 13 33 49.64   & $-$31 28 31.2   &    3.23    &    ext.   &   21.68   &   4.25   \\
$\#$12779     & 13 33 50.37   & $-$31 28 39.2   &   16.59    &     E     &   10994   &   1.50   \\
              &               &                 &            &           &           &          \\
J1333$-$3129  & 13 33 57.94   & $-$31 29 05.8   &    1.09    &  unres.   &    $-$    &   6.45   \\
              & 13 33 57.17   & $-$31 29 21.8   &   17.45    &     E     &    $-$    &   $-$    \\
              &               &                 &            &           &           &          \\
J1333$-$3158  & 13 33 58.72   & $-$31 58 11.9   &    1.00    &  unres.   &   21.43   &   1.78   \\
  $\#$12912   & 13 33 58.68   & $-$31 58 17.0   &   17.75    &     S     &   14689   &   1.75   \\
              &               &                 &            &           &           &          \\
\noalign{\smallskip}
\hline
\end{tabular}
\end{center}
\end{table*}
%
%
\setcounter{table}{3}
\begin{table*}
\caption[]{Optical Identifications. Continued}
\begin{center}
\begin{tabular}{lllrccc}
\hline\noalign{\smallskip}
Radio Name    & RA$_{J2000}$  & DEC$_{J2000}$ & $S_{1.4GHz}$ & Radio type &logP$_{1.4GHz}$  &  ${\cal R}$ \\
VLA$-$        &               &               &    (mJy)     &            & h$^{-2}$  (W Hz$^{-1}$) &             \\
Opt. Name     & RA$_{J2000}$  & DEC$_{J2000}$ &     $b_J$    & Opt.type   &        v        &  B--R       \\
              &               &               &              &            &   (km s$^{-1}$) &             \\  
\\
\noalign{\smallskip}
\hline\noalign{\smallskip}
J1334$-$3143  & 13 34 06.49   & $-$31 43 37.4   &    0.97     &  unres.   &    $-$     &   4.80   \\
MGP94 4345    & 13 34 06.5    & $-$31 43 39     &   18.48     &   $-$     &    $-$     &   $-$    \\
              &               &                 &             &           &            &          \\
J1334$-$3153  &  13 34 11.15  & $-$31 53 36.0   &    0.66     &  unres.   &    $-$     &   0.62   \\
              &  13 34 11.25  & $-$31 53 34.8   &   18.39     &     E     &    $-$     &   $-$    \\
              &               &                 &             &           &            &          \\
J1334$-$3149b & 13 34 17.71   & $-$31 49 07.8   &    1.21     &  unres.   &    $-$     &   1.96   \\
$\#$13013     & 13 34 18.16   & $-$31 49 08.0   &   17.25     &     S     &    $-$     &   $-$    \\
              &               &                 &             &           &            &          \\
J1334$-$3119b & 13 34 22.11   & $-$31 19 08.4   &   41.70     &  unres.   &    $-$     &   3.40   \\
              & 13 34 22.3    & $-$31 19 18     &   20.63     &    $-$    &    $-$     &   $-$    \\
              &               &                 &             &           &            &          \\
J1334$-$3137  & 13 34 22.82   & $-$31 37 08.3   &    0.74     &  unres.   &    $-$     &   0.99   \\
              & 13 34 23.00   & $-$31 37 10.0   &   21.12     &    $-$    &    $-$     &   $-$    \\
              &               &                 &             &           &            &          \\
J1334$-$3141a & 13 34 36.65   & $-$31 41 01.6   &    2.11     &  unres.   &    21.52   &   3.91   \\
$\#$13198     & 13 34 35.87   & $-$31 41 07.0   &   16.48     &     S     &    11357   &   $-$    \\
              &               &                 &             &           &            &          \\
J1334$-$3132  & 13 34 37.36   & $-$31 32 47.9   &   26.48     &   ext.    &    22.86   &  11.46 $\diamond$\\
$\#$13281     & 13 34 39.9    & $-$31 32 55     &   17.30     &     E     &    16490   &   1.80   \\     
              &               &                 &             &           &            &          \\
J1334$-$3126b & 13 34 40.57   & $-$31 26 32.4   &    2.71     &  unres.   &    21.82   &   3.88   \\
$\#$13277     & 13 34 40.76   & $-$31 26 43.4   &   17.09     &     E     &    14156   &   1.68   \\
              &               &                 &             &           &            &          \\
J1335$-$3139  & 13 35 02.97   & $-$31 39 11.6   &   16.71     &  unres.   &    22.66   &   3.09   \\
  $\#$13503   & 13 35 03.14   & $-$31 39 20.3   &   15.73     &     E     &    15077   &   2.07   \\
              &               &                 &             &           &            &          \\
J1335$-$3133  & 13 35 12.17   & $-$31 33 37.3   &    2.21     &  unres.   &    21.69   &   3.33   \\
$\#$13629     & 13 35 12.59   & $-$31 33 45.5   &   16.36     &     S     &    13673   &   1.50   \\
              &               &                 &             &           &            &          \\
J1335$-$3134  & 13 35 18.98   & $-$31 34 41.9   &    2.77     &  unres.   &    23.08   &   5.63   \\
$\#$13630     & 13 35 18.86   & $-$31 34 58.2   &   18.17     &   $-$     &    58050   &   $-$    \\ 
\noalign{\smallskip}
\hline
\end{tabular}
\begin{flushleft}
\vspace{0.5cm}
Notes to Table 4: $\diamond$ see comments in Appendix.
\end{flushleft}
\end{center}
\end{table*}
%
%

The morphological classification of the optical objects given in Table 4
was done by inspection of the DSS--2 images. The redshift 
information is taken from Bardelli et al. 
(\cite{bardelli94}, \cite{bardelli98b}) and Metcalfe et al. 
(\cite{metcalfe94}).

We found 68 identifications, corresponding to $\sim$40\% of 
our radio source sample. Among these, 33 (48\% of the identified sources)
are located at the redshift of the 
Shapley Concentration (velocity range $\sim$ 11000 $-$ 17500 km s$^{-1}$, 
Bardelli et al. \cite{bardelli98b}). 

\section{The radio emission in A3558--C}\label{sec:prop}

\subsection{General comments on the radio galaxies}

The sample of 33 optically identified radio sources in A3558--C 
includes 18 early--type 
galaxies and 14 spirals. The optical morphology of J1333$-$3124 is 
unavailable (see Table 4).

In Figure \ref{fig:spirali} and \ref{fig:ellittiche} we show the 1.4 GHz 
radio contours respectively of the late-- and early--type radio galaxies, 
overlaid on the DSS--2 optical frame. We note that the radio galaxy 
J1332--3146a is shown separately in Fig. \ref{fig:J1332-3146a}; the
radio halo at the centre of A3562, including the radio galaxy J1333--3141
is given in Fig. \ref{fig:estese}; and finally J1333--3124 is included
among the early--type galaxies (Fig. \ref{fig:ellittiche}).

%
\begin{figure*}
\centering
\includegraphics[angle=0,width=\hsize]{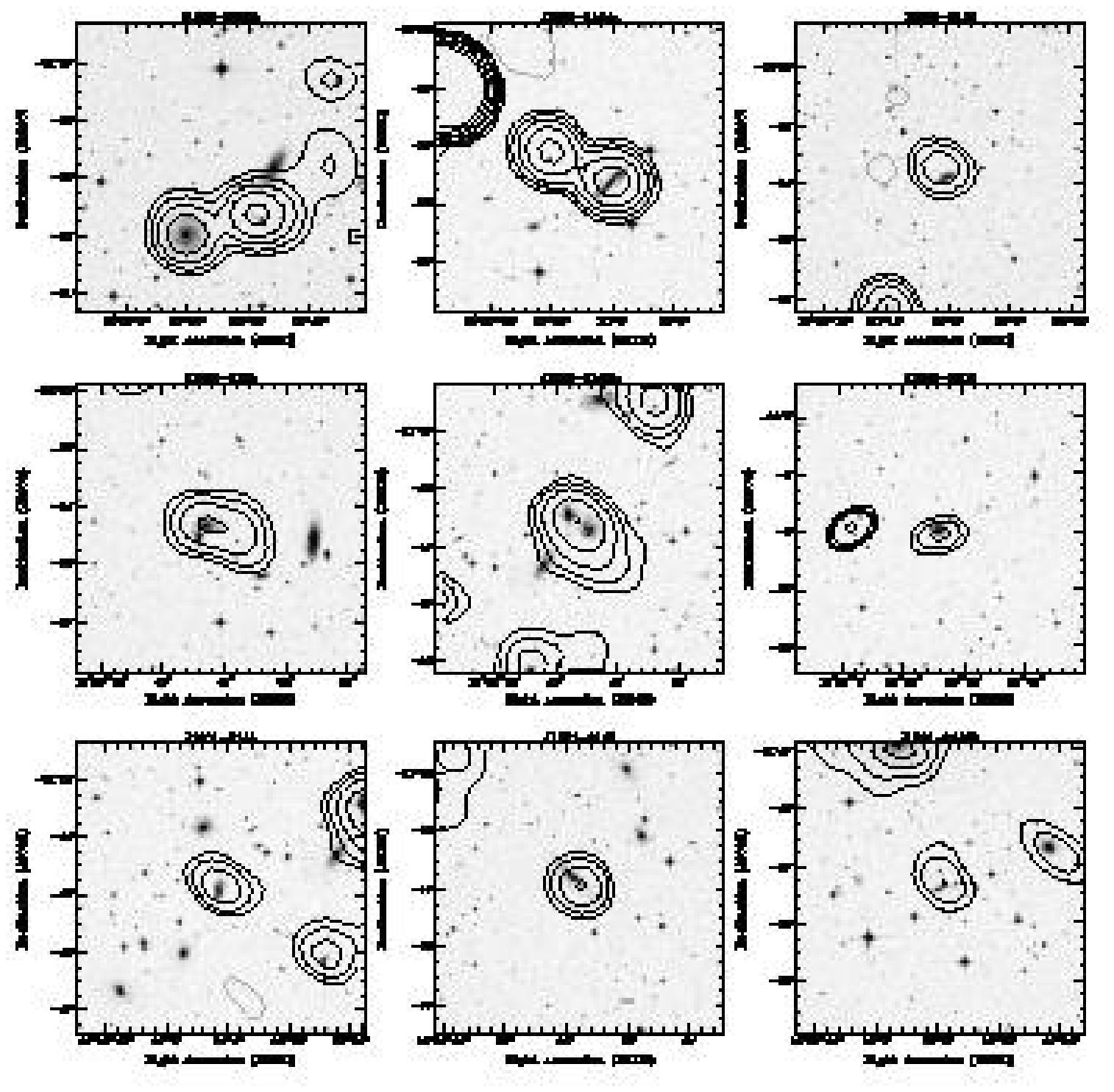}
\caption[]{1.4 GHz radio contours of the 14 late--type radio galaxies in
A3558--C overlaid on the DSS--2 optical frame. Radio contours are as follows.
For J1330$-$3209: --0.25, 0.25, 0.30, 0.35, 0.50, 1.00, 2.00 mJy b$^{-1}$; 
for J1332$-$3152a: --0.25, 0.25, 0.30, 0.50, 1.00, 2.00 mJy b$^{-1}$;
for J1335$-$3133: --0.50, 0.50, 0.75, 1.00, 1.50, 2.00 mJy b$^{-1}$.
for all remaining radio galaxies:
--0.25, 0.25 0.50, 1.00, 2.00 mJy b$^{-1}$; 
The restoring beam is $41.98^{\prime\prime} \times 35.13^{\prime\prime}$, 
p.a. $55.7^{\circ}$. } 
\label{fig:spirali}
\end{figure*}
%
%
\setcounter{figure}{1}
\begin{figure*}
\centering
\includegraphics[angle=0,width=\hsize]{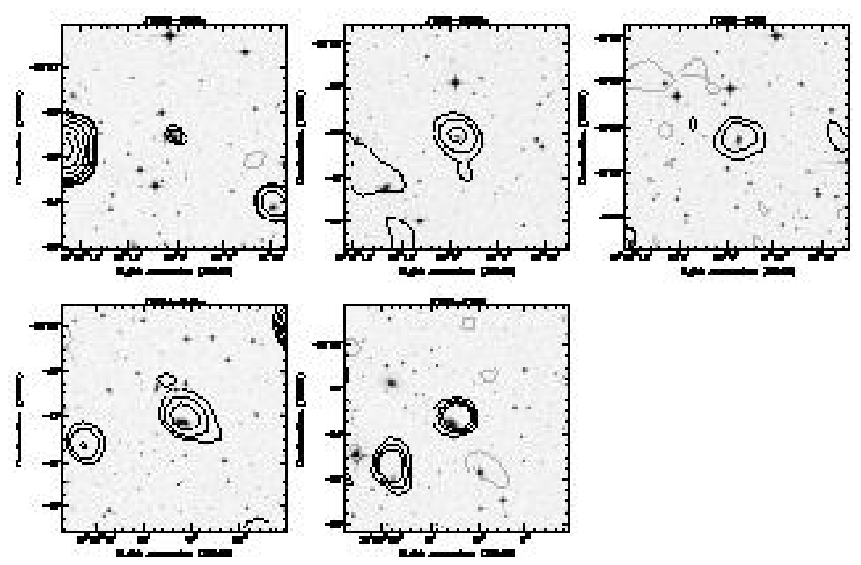}
\caption[]{ Continued.} 
\end{figure*}
%
%
\begin{figure*}
\centering
\includegraphics[angle=0,width=\hsize]{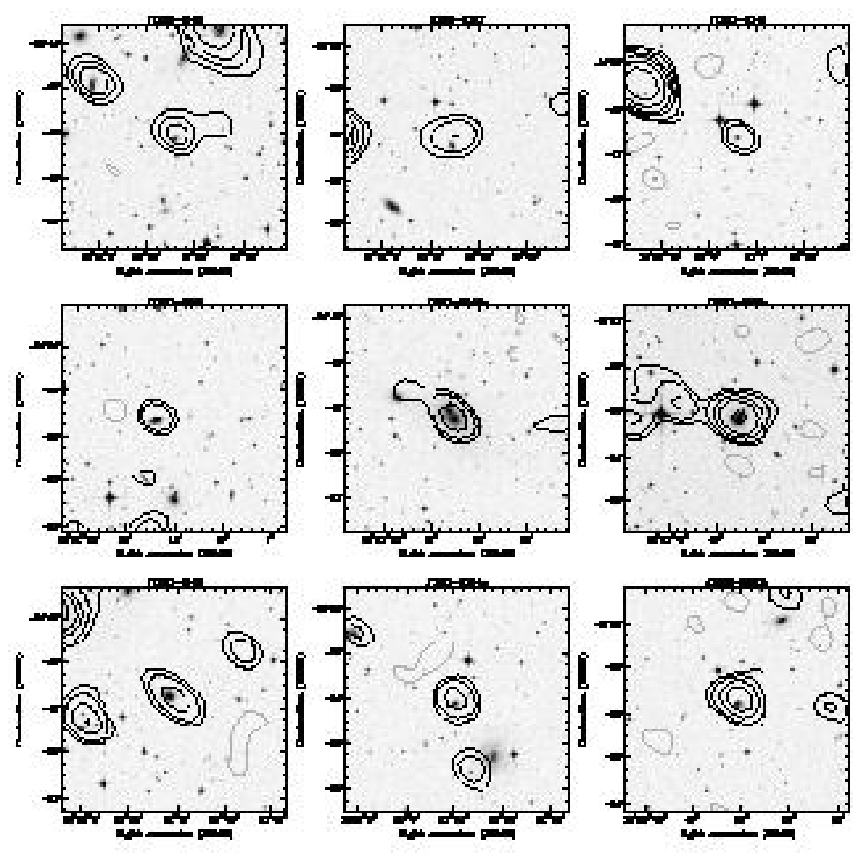}
\caption[]{1.4 GHz radio contours of the early--type radio galaxies in 
A3558--C overlaid on the DSS--2 optical frame. Radio contours are: --0.175, 0.175,
0.3, 0.6, 1, 2, 4, 8, 16 etc. mJy b$^{-1}$. For J1330$-$3146 and J1334$-$3126b 
the lowest contour is --0.3, 0.3 mJy b$^{-1}$. 
The restoring beam is 
$41.98^{\prime\prime} \times 35.13^{\prime\prime}$, p.a. $55.7^{\circ}$.}
\label{fig:ellittiche}
\end{figure*}
%
%
%
\setcounter{figure}{2}
\begin{figure*}
\centering
\includegraphics[angle=0,width=\hsize]{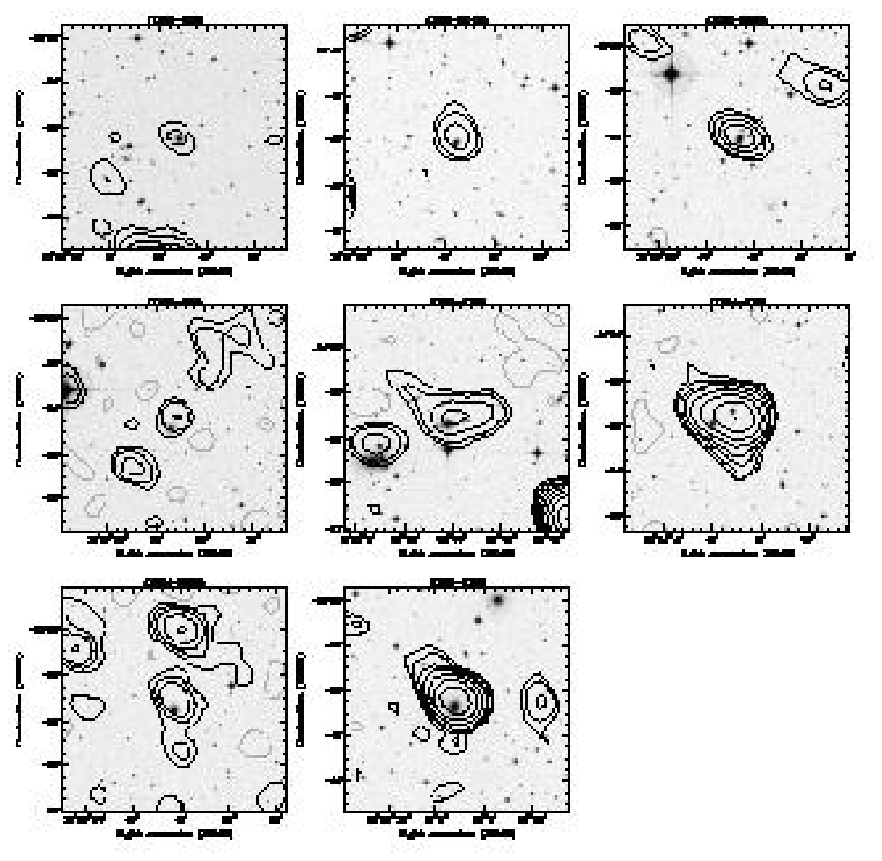}
\caption[]{Continued.} 
\end{figure*}
%
%
%
\begin{figure}
\centering
\includegraphics[angle=0,width=\hsize]{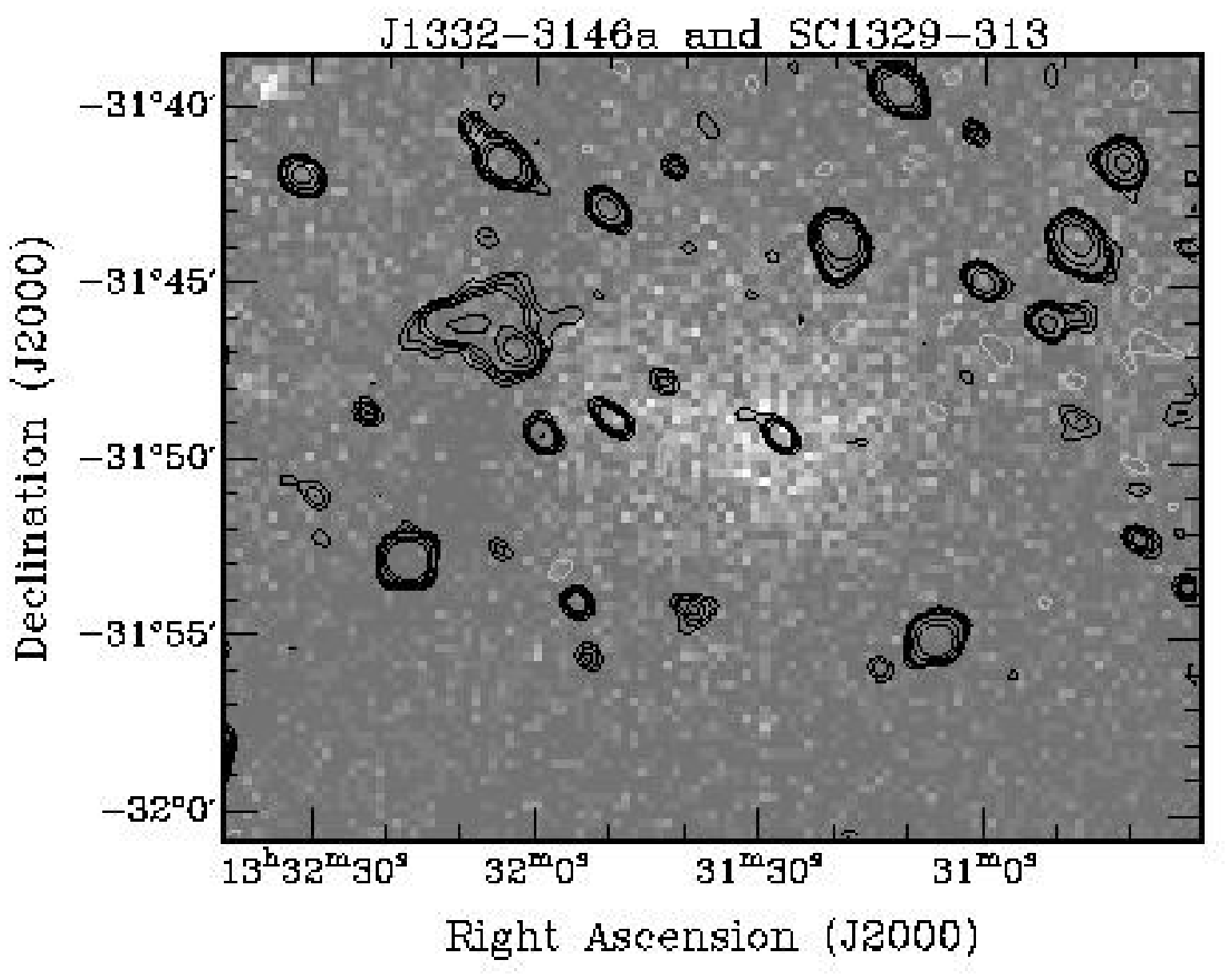}
\caption{1.4 GHz VLA radio emission from the group SC 1329--313 (grey scale)
superposed to the ROSAT X--ray emission. The extended radio galaxy 
J1332--3146a is located just outside the X--ray peak, in the 
North--East direction.
The FWHM of the restoring beam of the radio image is 
$41.98^{\prime\prime} \times 35.13^{\prime \prime}$, 
in p.a. 55.7$^{\circ}$.
The radio contours are --0.2, 0.2, 0.3, 0.4, 0.5, 1, 2 
mJy b$^{-1}$.}  
\label{fig:J1332-3146a}
\end{figure}
%
%
%
\begin{figure*}
\centering
\includegraphics{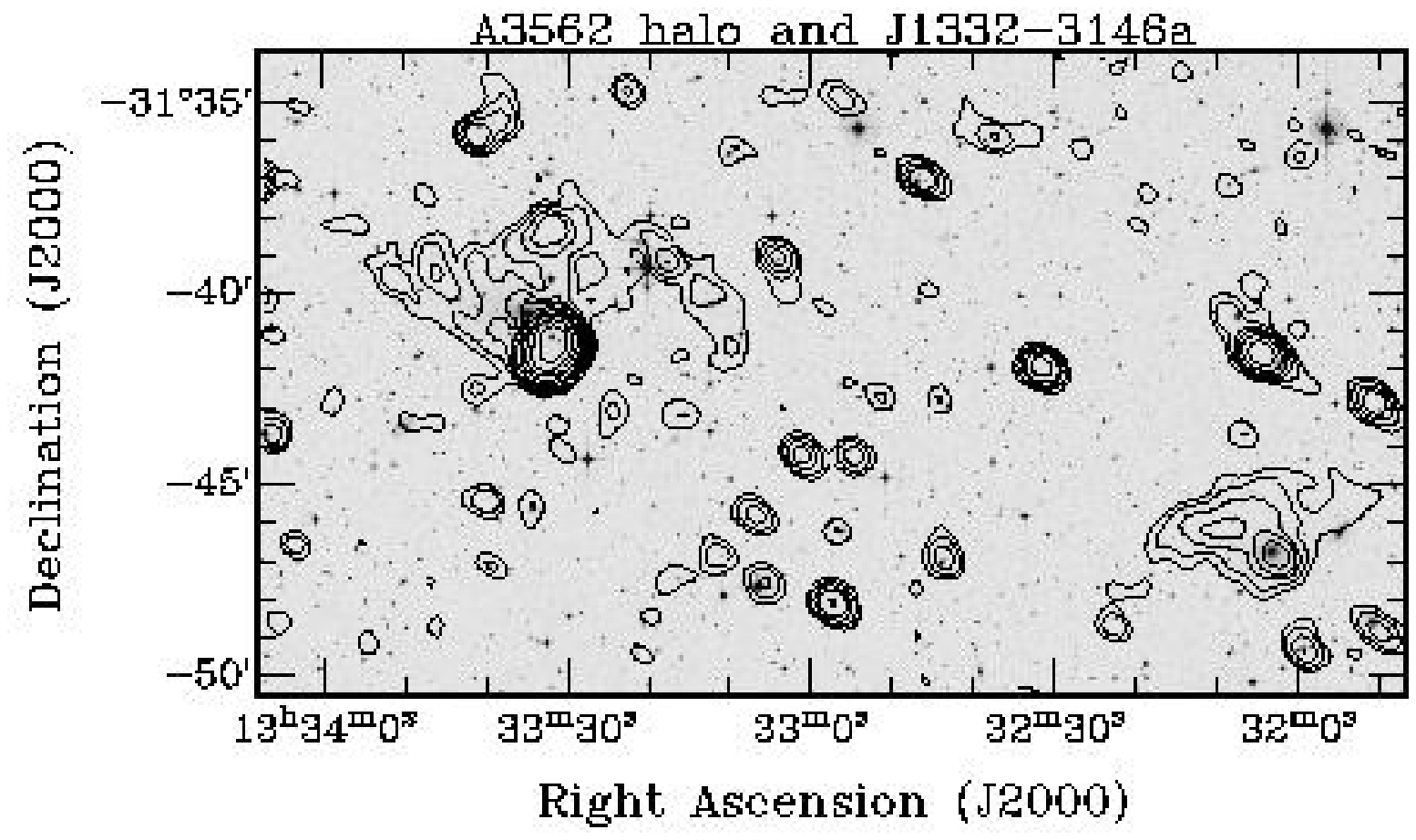}
\caption{1.4 GHz radio contours of the A3558--C region between the
A3562 central radio halo and the extended radio galaxy J1332--3146a,
overlaid on the optical DSS--2 red frame.
Contours are --0.15, 0.15, 0.3, 0.6, 1, 2, 4, 8, 16, 32, 64, 128, 250
mJy b$^{-1}$. The restoring beam in the image is 
$41.98^{\prime\prime} \times 35.13^{\prime\prime}$, in p.a. 55.7$^{\circ}$.}
\label{fig:estese}
\end{figure*}
%
%
%
\begin{figure}
\centering
\includegraphics[angle=0,width=\hsize]{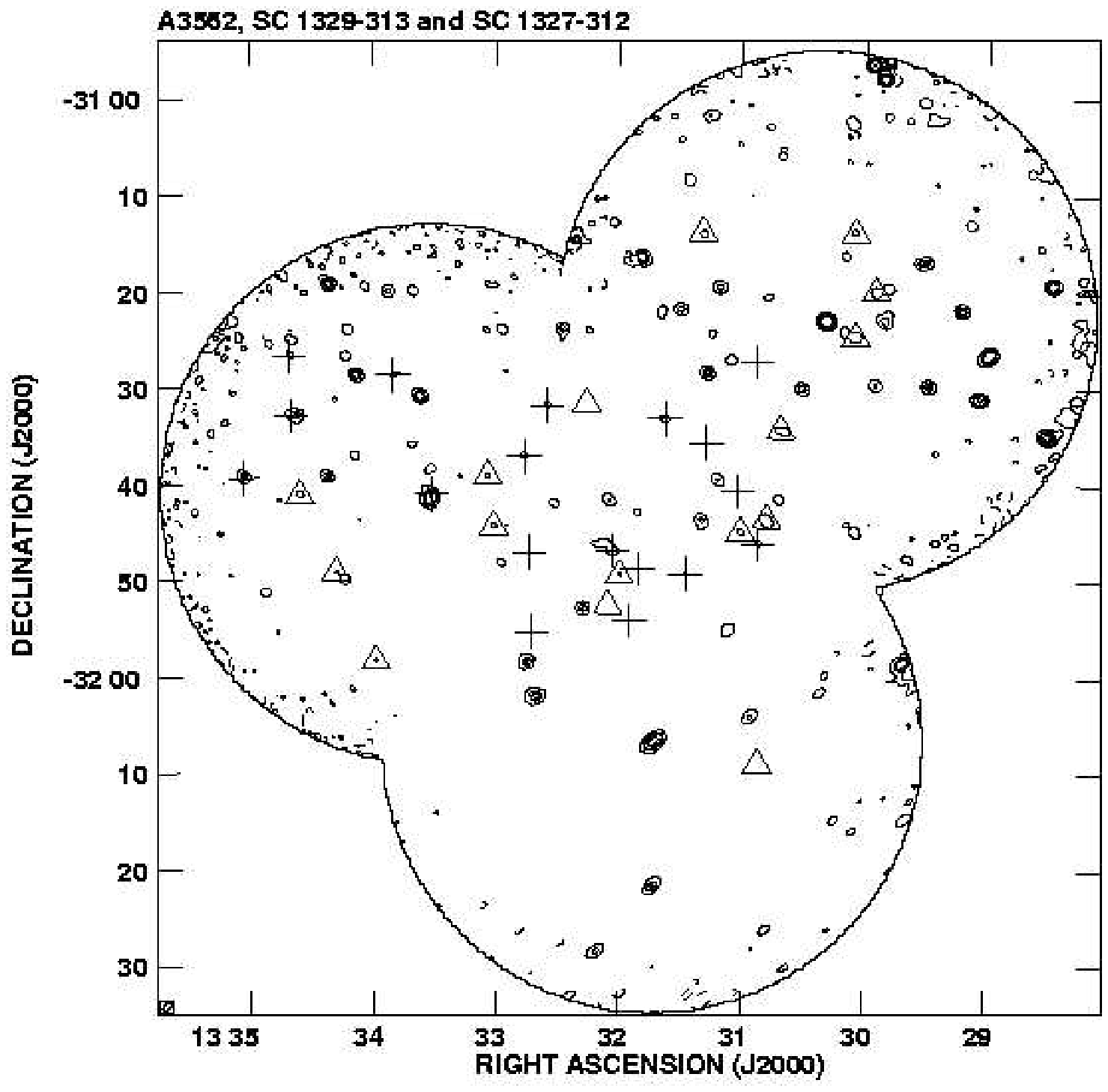}
\caption{Location of the Shapley Concentration optical counterparts overlaid 
on the 1.4 GHz 
radio contours of the A3562 and SC groups region. Triangles represent 
late--type galaxies and crosses stand for early--type galaxies. 
The radio contours are --0.85, 0.85, 6.80, 13.60 and 54.40 mJy b$^{-1}$.
The galaxy without optical classification was included among the early--type
galaxies.}  
\label{fig:mosaic}
\end{figure}
%
%
In Figure \ref{fig:mosaic} we show the location of the A3558--C 
early-- and late--type radio galaxies  listed in Table 4, overlaid on the 
1.4 GHz radio contours of the same region. No difference in
the bidimensional distribution of these two classes is obvious from the 
figure. 
\\
In search for possible segregation effects in the location of the
33 radio galaxies detected in this survey, 
we examined the velocity distribution of the optical counterparts, and compared 
it to that of all optical galaxies belonging to the A3558--C region covered
in our survey. The optical catalogue was  
extracted from the sample in Bardelli et al. (\cite{bardelli94}, \cite{bardelli98b}). 
The radio galaxies appear to be uniformly distributed over the Shapley supercluster
velocity range, with a peak of nine objects at $\sim$ 13000 km s$^{-1}$ 
(Fig. \ref{fig:histovel}). 
Six of them belong to the SC 1329-313 group and in particular to the T520 
subclump (Bardelli et al. \cite{bardelli98a}).

%
\begin{figure}
\centering
\includegraphics[angle=0,width=\hsize]{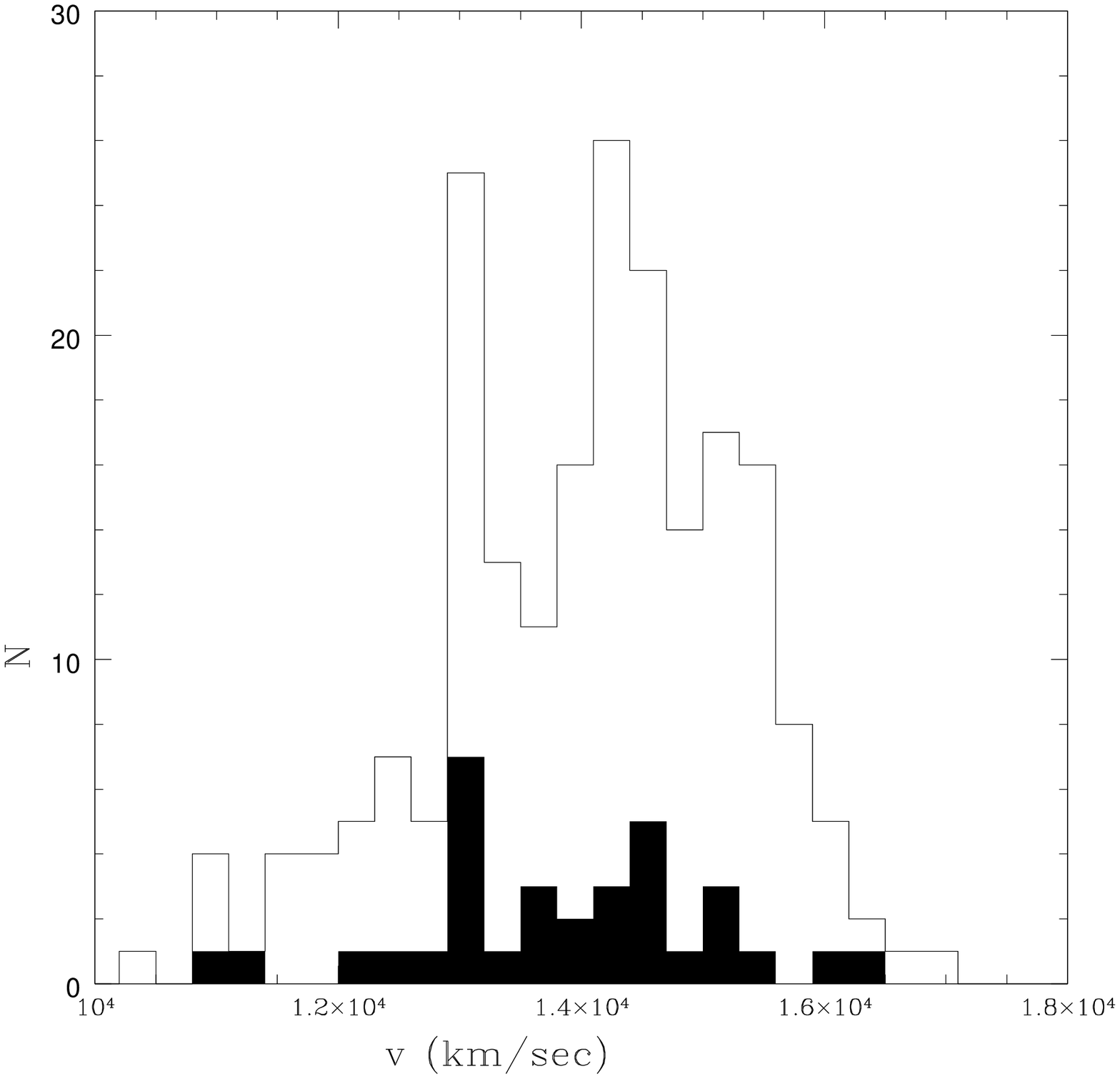}
\caption{Histogram of the velocity  distribution of the radio galaxies in the
A3562 and SC groups region given in Table 4 (black bins), compared to the velocity distribution 
of all galaxies in the Shapley Concentration. The width of each velocity bin is 300 km s$^{-1}$.}
\label{fig:histovel}
\end{figure}
%

\subsection{The diffuse radio galaxy J1332-3146a}

A remarkable feature of the radio emission in A3558--C 
is the very low brightness extended tail associated with the 
radio galaxy J1332$-$3146a. This source is 
identified with the cluster galaxy \#11744 ($b_J$ = 14.96 and  
velocity $v$ = 13107 km s$^{-1}$, see Table 4), the 
brightest galaxy in the group SC 1329--313. The projected 
angular size of the  radio emission is $\sim~6^{\prime}\times~4^{\prime}$, 
corresponding to $\sim~240\times~160$ kpc. The radio power is
logP$_{1.4GHz}$ (W Hz$^{-1}$) = 22.52  and its surface
brightness is $\sim$ 0.16 mJy b$^{-1}$.
\\
We note that the X--ray emission in this region is elongated in 
the direction of A3562, and compressed towards SC 1327--312, and 
J1332--3146a is located at the border of the X--ray emission of the 
SC 1329 group, as clear from Figure \ref{fig:J1332-3146a}.

The presence of extended emission associated with J1332$-$3146a was suggested 
by inspection of 843 MHz Molonglo Observatory Synthes Telescope (MOST)
observations and of the NVSS 1.4 GHz image, where the source has a projected 
angular size of $\sim~4^{\prime}\times~2^{\prime}$ (V2000). 
The overall morphology in our image (Figure \ref{fig:J1332-3146a}) 
is in very good agreement with the NVSS image.

J1332$-$3146a was detected as point--like with the ATCA both at 22 cm 
and 13 cm, at the resolution of a few arcseconds, and 
the derived spectral index is $\alpha_{13}^{22}$ = 0.81 (V2000). 
We note that the source was labelled J1332$-$3146 in V2000.
The compact morphology and steep spectrum at high resolution seem to rule 
out any connection between the low brightness extended emission and an 
active nucleus. In other words, there is no indication that the extended 
emission imaged with the observations presented here is a tail of a 
head--tail or wide--angle tail radio source.

In Figure \ref{fig:estese} the radio contours of the A3558--C region
between the radio halo at the centre of A3562 and J1332--3146a are superposed 
to the DSS--2 optical frame. Inspection of the radio emission in this
region suggests some important considerations. In particular,
{\it (a)} the extension of J1332--3146a points towards the radio halo at 
the centre of A3562; {\it (b)} the radio halo is characterised by a 
filament pointing South--West, towards J1332--3146a;
{\it (c)} positive residuals of radio emission are clearly 
visible in the region between the radio halo and J1332--3146a. 
Excluding the strongest point sources (with flux density S$> 5\sigma$),
these residuals account for several mJy.

It is possible that the extended emission in J1332--3146a is {\it (i)}  
part of a very low brightness cluster scale bridge of emission,
extending from the centre of A3562 to the smaller group SC 1329--313,
undetectable with the radio instrumentation presently available;
{\it ii)} the result of reacceleration processes and can therefore be 
interpreted as {\it revived} radio
emission. We remind here that J1332$-$3146a lies in a region where
a merger shock is expected, on the basis of numerical simulations
and of the properties of the X--ray emission.

\subsection{The population of faint radio galaxies in A3558--C. 
Enhanced radio activity due to a cluster merger?}\label{sec:faint}

An indicator of starburst activity is the presence 
of radio emission at power levels of the order 
P$_{1.4~GHz}$ \ltsim $10^{23}$h$^{-2}$ W Hz$^{-1}$
(see Condon et al. 2002 for a recent discussion).
The majority of the A3558--C radio galaxies in our list has radio power 
below this threshold. In particular, 12 out of 18 radio emitting ellipticals
have logP$_{1.4~GHz}$ (W Hz$^{-1}$) $<$ 21.78, the lower limit of the
radio luminosity function for AGNs derived by LO96 (scaled to the cosmology
adopted in our paper) discussed in Section 7.1. Among the
spiral galaxies with associated radio emission, they all have 
logP$_{1.4~GHz}$ (W Hz$^{-1}$) $\le$ 22.25.
Including the radio galaxy J1333--3124, whose optical counterpart
has no morphological classification, the total number of candidate 
starburst radio galaxies in our sample is therefore 26.

A detailed optical study of the spectral properties of the galaxies 
in A3558$-$C shows that the region between A3562 and the 
SC groups is populated by a very large fraction of blue galaxies (Bardelli et 
al. \cite{bardelli98b}).
To quantify this effect we considered the (B--R) colours of
the photometric sample of optical galaxies in this region of Metcalfe et al. 
(\cite{metcalfe94}): from a b$_J$ vs (B--R) plot we found that the red sequence 
is at (B--R)~$~\sim~$~2 and we set (B--R) $<$ 1.7 as the
limit to define a blue object.
We cross--correlated our sample of faint radio galaxies with the
sample of Metcalfe et al. (\cite{metcalfe94}) and found that 
4 spirals and 7 ellipticals are blue.
Beyond that, we note that the (B--R) values for all the radio galaxies 
in A3558--C are lower than the average value for A3558--C [$<$(B--R)$>~\sim~$2], 
with few very extreme cases (see Table 4). In conclusion, for half 
of the faint radio sources in the region under study, indication of starburst 
radio emission is supported by photometric information. Assuming that there is no
contribution from an active nucleus, the star formation rates (SFR)
deduced on the basis of the radio emission are in the range 
SFR$\sim~1.2 \div 7.1$ M$_{\odot}$ yr$^{-1}$. This estimate
was done using the empirical relation between SFR and radio luminosity
in Yun et al. (2001), after scaling our radio luminosity
to the cosmology used in their paper.

We compared our results on A3558--C with those obtained by MO03
for A2255, whose radio emission
was interpreted in terms of increased radio AGN activity and star
formation as consequence of the cluster merger.
\\
For a proper comparison we needed to apply some transformations
and corrections to our data. In particular:

\parn (i) first of all, for our COSMOS $b_J$ magnitudes, we considered a correction 
according to Lumsden et al. (\cite{lumsden97}) for
a ``saturation'' effect which occurs for bright objects due to
the lack of dynamic range within the measuring machine; then
we assumed $b_J$ \simeq $m_B$ (as found by Bardelli et al. \cite{bardelli00})
and applied a correction for the galactic absorption of A$_B$ = 0.24 (Schlegel et al.
\cite{schlegel98}). We converted our magnitudes to the m$_{Rc}$ magnitudes using
the (B--R$_c$) colours in Fukugita et al. (\cite{fukugita95}): we adopted 
(B--R$_c$) = 1.48 for early-type galaxies (average value between E and S0)
and (B--R$_c$) = 1.00 for late-type galaxies (average value between spirals and Irr).
Finally we computed the $M_R$ magnitudes for our sample adopting the MO03 cosmology.

\parn (ii) We rescaled the radio powers of our A3558--C radio galaxies
to the MO03 cosmology,  
and selected only those sources with logP$_{1.4~GHz}$ (W Hz$^{-1}$) $\ge$ 21.84.

\parn (iii) We extracted a normalization sample from the COSMOS catalogue 
containing all
galaxies within the MO03 limit $M_R \le -20$ in the region of A3558--C covered 
by our observations.
We applied the same magnitude corrections described in point (i). 
In this case we adopted
an average (B--R$_c$) = 1.36 with the assumption of a morphological mix of
73\% ellipticals and 27\% spirals in this region. From the number
counts of this sample we subtracted the contribution of field galaxies 
determined from the ESP survey (Vettolani et al. 1997) which was obtained 
from the COSMOS catalogue. Also for these
objects we considered the same magnitude corrections as in point (i) with A$_B$ = 0
(the ESP survey is at the galactic pole) and
(B--R$_c$) = 1.15  which corresponds to a morphological mix for the field of
30\% early--type and 70\% late--type galaxies.

Following MO03 we divided our optical sample in three magnitude bins, 
i.e. ``faint'' ($-21 < M_R \le$ -20), ``intermediate'' ($-22 < M_R \le -21$)
and ``bright'' ($M_R \le -22$), and computed the fraction
$N_{rg} \over N_{gal}$ in each luminosity bin, including all optical
morphological types. As in their paper,
$N_{gal}$ and $N_{rg}$ are respectively the total number of 
optical galaxies and the total number of radio galaxies in each bin.
The results of our analysis are given in Table 5 where we also
report the numbers for A2255 and the mean and the dispersion
for all the other clusters considered by MO03 in the 2Mpc case. 
We note that the fractions and errors associated with the MO03
clusters were computed by us using the numbers given in their paper.

%
\setcounter{table}{4}
\begin{table*}
\label{tab:mo03}
\caption[]{Radio--optical galaxy counts}
\begin{center}
 \begin{tabular}{l|rrr|rr}
\hline\noalign{\smallskip}

  Mag Interval    &  $N_{rg}$   & $N_{gal}$     & \%              & \%              &       \%         \\ 
                  &\multicolumn{3}{c}{   (A3558--C)}              & (A2255)         &   Other clusters \\
\hline 

$-21 < M_R \le -20$ &   3       &   88.4      &  3.4\%$\pm$2.0    &  7.2\%$\pm$2.7  &   1.6\%$\pm$1.4  \\
$-22 < M_R \le -21$ &   9       &   37.8      & 23.8\%$\pm$7.9    & 15.3\%$\pm$4.8  &  20.3\%$\pm$12.8 \\
$M_R \le -22$       &   6       &   13.7      & 43.8\%$\pm$17.9   & 56.5\%$\pm$21.3 &  23.3\%$\pm$23.1 \\
\hline
\end{tabular}
\end{center}
\end{table*}
%
Not surprisingly, the ratio $N_{rg} \over N_{gal}$ increases
going from the faint to the bright optical bin both in 
A3558--C and in A2255. If we take into account the associated errors, 
there is only marginal evidence of enhanced fraction of radio emitting
galaxies in A3558--C and in A2255 compared to the other clusters
considered in MO03.

\section{Statistical analysis of A3558--C: radio/optical 
properties}\label{sec:RLF}

The radio luminosity function (RLF) is a powerful statistical 
tool to investigate the radio properties of a galaxy population. 
In order to understand if the ongoing major merger in 
A3558--C has significant effect on the
radio emission of the cluster galaxy population, it is important to
compare the radio luminosity function (RLF) for galaxies
in this merging environment with the mean RLF for galaxies in normal 
clusters and in the field.
\\
With this aim we computed the RLF for the  
radio galaxies in the A3558--C and compared our results with those
obtained by LO96 and by Gavazzi \& Boselli (\cite{gavazzi99}, 
hereinafter GB99) respectively for early-- and late--type galaxies. 

For the comparison we adopted the cosmology used by these
authors and scaled the radio powers of our sample. We corrected 
our $b_J$ magnitudes as described in point (i) of Section \ref{sec:faint}.

\subsection{The radio luminosity function of AGNs}

We computed the number of radio galaxies expected in the A3558--C
region on the basis of the ``universal'' RLF derived by LO96 
and compared it to the number of objects actually detected in
our survey. 

The analysis carried out by LO96 
includes radio galaxies with logP$_{1.4~GHz}$(W Hz$^{-1}$) $\ge$ 22.03
and optical counterparts brighter than $M_R=-20.5$. This magnitude limit 
corresponds to $m_R=16.05$ at the distance of the Shapley supercluster.

We estimated that the total number of early--type galaxies in
the A3558--C region located at the distance of the Shapley Concentration
is 65 (see V2000 for further details). On the basis of LO96, 
the fraction of early$-$type galaxies with $M_R\le -20.5$ expected to be 
radio loud at a power level logP$_{1.4~GHz}$(W Hz$^{-1}$) $\ge$ 22.03  
is 6 ($\sim$ 9.2\%), in good agreement with the 
seven radio galaxies in our sample matching the LO96 constraints.

Conversely, V2000 found that the RLF for early--type galaxies of the whole
A3558 complex, including the cluster A3558 and the westernmost A3556, is 
significantly lower than the ``universal'' RLF by LO96.
These results suggest that the effect of cluster merger on the
AGN--type radio emission may be a very complex phenomenon, possibly
depending on the stage of the merger. This issue will be further
discussed in Section 8.

\subsection{The radio luminosity function of late--type galaxies}

The role of environment on the radio emission of spiral galaxies
has been investigated in a number of papers.
Jaffe \& Perola (\cite{jaffe76}) found that radio emitting spirals
in the Coma cluster have a radio excess compared to field galaxies. 
Gavazzi \& Jaffe 
(\cite{gavazzi86}) confirmed these results by comparing the 
RLF of late--type galaxies within and outside rich clusters. 
GB99 computed the RLF of spiral galaxies, in terms of 
radio--optical flux density ratio, in nearby different 
environments (rich and poor clusters, and the field),  
and found that late--type galaxies in rich clusters 
develop radio sources more frequently than galaxies in poor 
clusters and in the field. They also found a correlation 
between the radio excess and the velocity deviations with 
respect to the cluster average velocity. 
GB99 suggested that these results are coherent with a ram
pressure scenario: galaxies in fast motion through the intracluster
medium experience enough dynamical pressure to compress their 
magnetic field on the up-stream side, form a tail-like radio
structure on the down-stream side and produce a net enhancement
of the radio continuum activity.
  
We addressed the question whether the late--type galaxy population
in the merging environment of the A3558--C region 
behaves like those in rich clusters, or if it shows even more
enhanced radio excess.
To this aim we compared the RLF of late--type galaxies in this region 
with the results obtained by GB99. We 
extracted a subsample of 209 objects from the GB99 sample, with
Zwicky magnitude $m_z$ $<$ 15.7 and flux density S$_{1.4GHz}$ 
$\ge$ 2.25 mJy, corresponding to the 5$\sigma$ level of the NVSS 
survey (radio data in GB99 are taken mostly from this survey). 
This flux density limit implies a radio power limit of 
logP$_{1.4~GHz}$(W Hz$^{-1}$) = 20.74 at the distance of the Cancer cluster,
the nearest in their sample ($z=0.015$).

For a proper comparison we computed the $b_J$ magnitude limit corresponding
to the $m_z$ limit in GB99. First we converted $m_z$ in $m_B$ according to 
the relationship $m_B=m_z - 0.35$ (Gazta\~naga \& Dalton, 2000), assuming
$b_J$ \simeq $m_B$ (Section \ref{sec:faint}). Then we applied a distance 
modulus of 1.5 mag and a 0.24 magnitude correction for
the galactic absorption (Schlegel et al. \cite{schlegel98}). We obtained 
a limit of $b_J^{\prime}$ = 17.09, where $b_J^{\prime}$ is the magnitude 
corrected as descibed above.

For each late--type radio galaxy in A3558--C we applied a correction 
for the internal extintion. This latter was determined according to Gavazzi \&
Boselli (\cite{gavazzi96}) using an average value of 0.60 for the D$_B$(type) 
coefficient, which corresponds to Sb galaxies in the B band. 

We selected, at the same radio power limit as GB99, a sample of spiral galaxies 
in A3558--C with $b_J^{\prime} \le$ 17.09. 
The resulting sample includes 13 objects and is presented in Table 6. 
We remind that the  radio powers in Table 6 are scaled to the cosmology adopted 
by GB99.

\begin{figure*}
\centering
\includegraphics[angle=0,width=0.90\hsize]{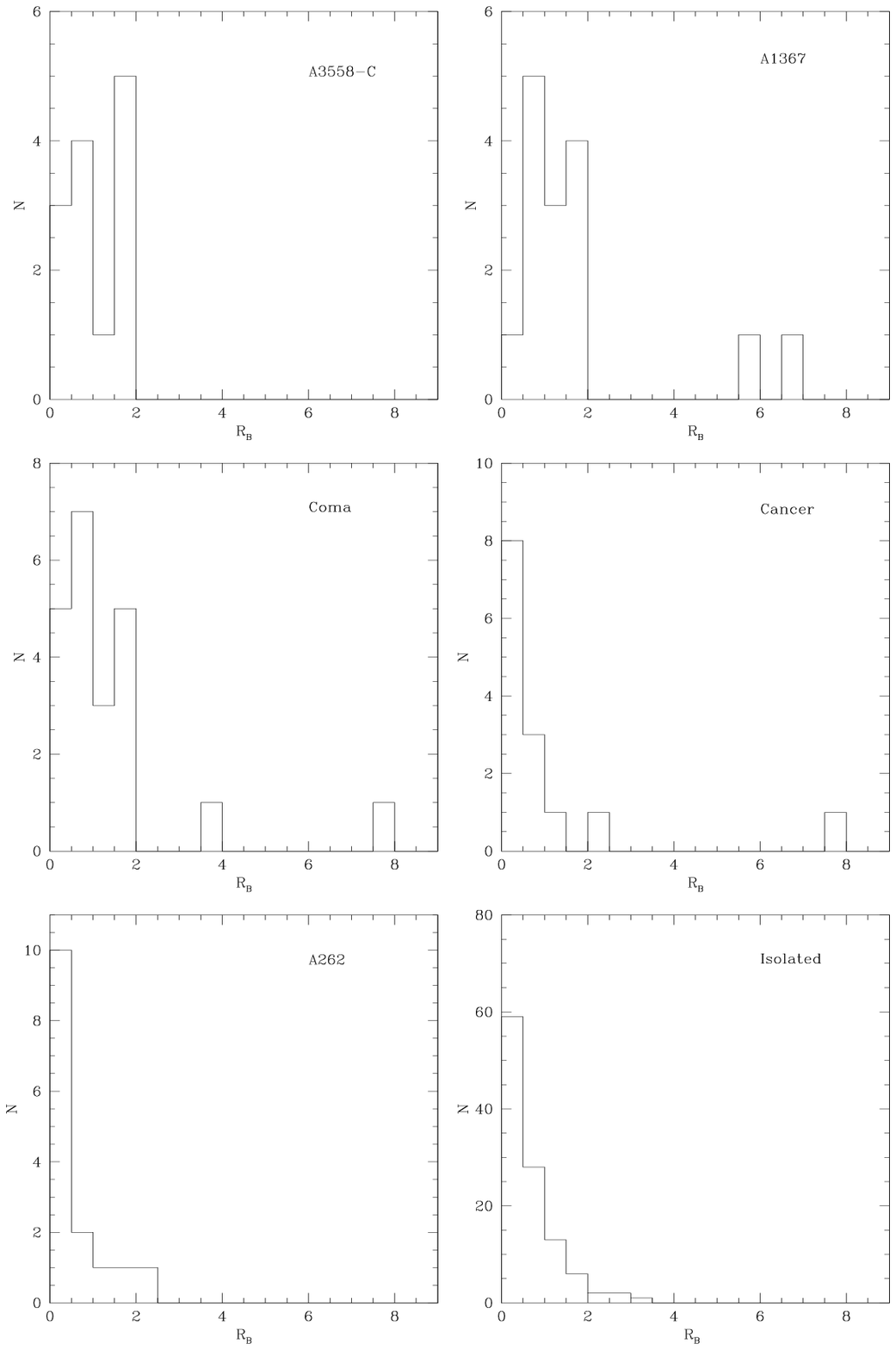}
\caption{Distributions of the R$_B$ radio/optical ratios for the late--type
galaxies in A3558--C, A1367, Coma, Cancer, A262 and for isolated
galaxies in the Coma supercluster.}
\label{fig:RBlate}
\end{figure*}

%
%
\setcounter{table}{5}
\begin{table*}
\label{tab:late}
\caption[]{Sample of late$-$type radio galaxies used for the RLF}
\begin{center}
 \begin{tabular}{lccccc}
\hline\noalign{\smallskip}

  Radio       &  Optical      & $b^{\prime}_J$ & S$_{1.4GHz}$ & logP$_{1.4GHz}$ & R$_B$ \\ 
  name        &  name       &                &      mJy     &    W Hz$^{-1}$  &       \\
\hline 
J1329$-$3119b &  \#10174    &   14.39        &    4.53      &      22.07      & 0.58  \\
J1330$-$3124a &  \#10318    &   14.52        &    6.89      &      22.20      & 0.99  \\
J1330$-$3113  &  \#10313    &   16.42        &    1.79      &      21.76      & 1.49  \\
J1330$-$3134  &  \#10725    &   15.66        &    3.86      &      22.06      & 1.60  \\
J1330$-$3143b &  \#10869    &   14.96        &    7.77      &      22.25      & 1.69  \\
J1330$-$3209  &  \#10888    &   15.91        &    0.73      &      21.17      & 0.38  \\
J1331$-$3144  &  \#11000    &   16.20        &    2.55      &      21.67      & 1.73  \\
J1331$-$3113  &  \#11243    &   14.88        &    2.38      &      21.83      & 0.48  \\
J1331$-$3149b &  \#11751    &   16.61        &    1.82      &      21.58      & 1.81  \\
J1332$-$3152a &  \#11753    &   15.67        &    0.49      &      21.04      & 0.20  \\
J1333$-$3158  &  \#12912    &   17.07        &    1.00      &      21.52      & 1.52  \\
J1334$-$3141a &  \#13198    &   15.56        &    2.11      &      21.52      & 0.80  \\
J1335$-$3133  &  \#13629    &   15.52        &    2.21      &      21.69      & 0.80  \\
\hline
\end{tabular}
\end{center}
\end{table*}
%
%
For each radio galaxy of our sample we computed the radio/optical ratio, 
defined in GB99 as 
R$_B$=S$_{1.4GHz}$ / k $\times$ 10$^{-0.4 \times b_J^{\prime}}$,
where k = 4.44 $\times$ 10$^6$ is the factor appropriate to transform 
in mJy the $b_J$ magnitude. We note that this quantity
is independent of distance. According to Gavazzi et al. (\cite{gavazzi96})
R$_B$ gives the ratio of the radio emission per unit light emitted 
by the relatively young stellar population. 
We compared the distribution of R$_B$ ratios for the late--type
galaxy sample in A3558--C to the subsample 
extracted from GB99 (Figure \ref{fig:RBlate}).
%
%
%
%
We point out that the R$_B$ distribution of our sample
does not change significantly if we use coefficients D$_B$(type) 
corresponding to morphological types different from Sb.  

The limited number of galaxies in A3558--C does not allow firm
conclusions, however some considerations can be made.
The most remarkable features in Fig. \ref{fig:RBlate} are the 
peaked distribution (in the first bin) for the galaxies in poor 
environments, and the presence of galaxies (though a limited
number) with a strong excess of radio emission both in rich and
poor environments.
\\
Considering poissonian errors in each bin, the distribution of 
radio late--type galaxies in A3558--C is substantially flat up to
R$_B$ = 2, similar to the distributions in Coma and A1367.
We estimated the similarity between A3558--C and Coma
applying a Kolmogoroff--Smirnov (KS) test to the two distributions and
found that the probability that they are the same distribution is
$\sim$92.4\%. A KS test applied to A3558--C and the isolated galaxies
gives a probability of only $\sim$3.6\%.

The late--type galaxies with R$_B$ \gtsim 4 are missing in A3558--C 
(respectively 2/15 and 2/22 for A1367 and 
Coma), however this could be due to lack of statistics, since 
only 13 spirals in A3558--C are included in the analysis.

To summarise, the statistical properties of the radio emission from 
spiral galaxies in A3558--C are similar to those in rich
and dynamically evolved environments.

\section{Discussion and Conclusions}\label{sec:disc}

In this paper we presented deep 1.4 GHz VLA observations of
the major cluster merger in the A3558 complex. The area under study,
A3558--C, defined as the region between the centres of the two Abell 
clusters A3558 and A3562, including also the two groups SC 1329--213
and SC 1327--312, is thought to have recently experienced
a major merger between two massive clusters. What we see
now is expected to be the result of the first core--core encounter. 
The observational properties in this region, together with numerical 
simulations, suggest that A3558 is the main cluster, while the
whole chain, beyond A3558 itself, is the remains of the colliding 
cluster (Bardelli et al. 2002). 
\\
A picture is emerging, in which radio AGN and starburst activity, 
radio relics and halos, or the lack thereof, are signature of cluster 
mergers at different stages. In particular, Venturi et al. (2002) proposed 
an evolutionary merger sequence to account for the diversity in the radio 
properties of the three main merging cluster systems in the core of the
Shapley Concentration; Kempner \& Sarazin (2001) postulated that
radio halos and relics may form at different times during
mergers; MO03 explained the different fraction
of radio emitting galaxies (starburst and AGNs) in a cluster sample
as due to different merger stages.
With this study we have a unique opportunity to connect the observed 
properties in the radio band with a well defined cluster merger
stage. This is of crucial importance for a better understanding of
the complex effects of cluster mergers, with particular emphasis 
to the stage of the merger. 
\\
The most relevant results of our analysis can be summarised as
follows:

\begin{itemize}

\item[(i)] a faint radio halo is found at the centre of A3562, whose
properties are consistent with the idea that it is a young source 
at the beginning of the reacceleration phase (V2003)
induced by a recent merger event;

\item[(ii)] the origin of the extended emission in the radio galaxy 
J1332--3146a, associated with the dominant galaxy in the group SC 1329--313, 
is unclear. No radio jets are present in the nuclear region of J1332--3146a, 
and the nuclear radio
component has steep spectrum. It is possible that this extended emission
has actually cluster origins (see Section 6.2), being either 
(a) a ``bright'' area of a very low brightness bridge of radio emission, 
connecting A3562 and SC 1329--313, or (b) a ``revived'' radio emission
region, where pre--existing old electrons were reaccelerated;

\item[(iii)] a large number of radio sources associated with 
A3558--C galaxies was found, i.e. 33 objects, most of them with
radio emission at low power levels. Our analysis (see Section 6.3) 
suggests that 26/33 radio galaxies are candidate starbursts.
Among them, 11 also show a blue excess;

\item[(iv)] the total number of radio AGNs detected in this region is
consistent with the expectations from the RLF of LO96, suggesting
that the cluster merger has not affected the probability of an
early--type galaxy to develop a nuclear radio source; 

\item[(v)] the distribution of the radio/optical ratios for
the spirals in A3558--C is similar to what is found in rich
and evolved environments.

\end{itemize}

The main question is if and how this wealth of observables is 
connected to the cluster merger in this region.
Beyond the noticeable finding of the radio halo in the centre
of A3562 (V2003), the radio properties of the Shapley 
galaxies in this region may contain important pieces of information.

V2000 showed that 
the RLF for early type galaxies in the whole A3558--C shows a deficit 
of radio galaxies 
compared to the ``universal'' RLF presented in LO96 (V2000),
over the whole power range. 
On the other hand, the RLF for AGNs presented here matches the 
expectations of LO96; furthermore the outskirts of the chain (A3556 and 
A3562) contain the largest fraction of Shapley radio galaxies (see also 
Section 2). This suggests that A3558 itself, the most
massive cluster in the chain, is the main responsible for the lack 
of radio sources in the RLF found in V2000. A possible explanation
is the key role of A3558 in the merger, i.e. it is experiencing 
the most dramatic effects of the merger, being the result of the
interaction of core regions of the two colliding clusters. 
If our interpretation is correct, then the role of cluster merger on 
the radio emission from AGNs may be many--fold, depending on the 
age and strength of the merger.
\\
Optical photometric information on the faint population of radio
galaxies in A3558--C confirms that at least 50\% of these objects
are most likely starburst candidates. 
However, a radio/optical analysis carried out following MO03 provides
only weak evidence that the fraction of radio emitting galaxies
in A3558--C is higher than in non--merging environments.
%
%
\\
The statistical results on the late--type galaxies show that 
the radio emission in the A3558--C spirals is similar to 
those in rich and dynamically evolved clusters. This suggests
that the radio emission in spiral galaxies may be one of the 
first ``parameters'' to react to a cluster merger event.

In conclusion, V2003 showed that the radio halo at the centre
of A3562 is consistent with a reacceleration phase which started
$\sim$ a few $10^8$ years ago. This ongoing merger is therefore
{\it advanced}, in the sense that the core--core encounter has
already taken place, but it is still young if compared to the
total duration expected for a cluster merger, i.e. 10$^9$ yr.
Our study suggests that on this timescale, the effect of cluster 
merger on the radio emission from cluster galaxies is many--fold.
In particular, we found only marginal evidence of enhanced radio 
emission of starburst origin, but we found significant enhanced 
radio emission from spirals. 
We argue that the role of cluster merger on the nuclear activity
in early--type galaxies is a complex phenomenon, since observational 
evidence in the whole A3558 cluster chain clearly shows that 
the deficit in radio galaxies found in V2000 is entirely
due to the cluster A3558, which has experienced the most violent
consequences of the merger.

\begin{acknowledgements}

This work has been partially supported by the Italian Space Agency grants
ASI-I-R-105-00, ASI-I-R-037-01 and ASI-I-R-063-02, and by the Italian 
Ministery (MIUR) grant COFIN2001 ``Clusters and groups of galaxies: the 
interplay between dark and baryonic matter". 
NRAO is a facility of the National Science Foundation, operated under
cooperative agreement by Associated Universities, Inc.
\\
This work has made use of the NASA/IPAC Extragalactic Database NED
which is operated by the JPL, California Institute of Technology, 
under contract with the National Aeronautics and Space Administration.

\end{acknowledgements}

\appendix

\section{Notes to optical identifications}

\parn $-$ J1329$-$3122a : this radio source is the blend of two sources 
in the ATCA 22 cm catalogue: the pointlike J1329$-$3122, associated with 
a 18.54--mag galaxy, and the FRII J1329$-$3123, identified with the elliptical 
cluster galaxy \#10178 ($b_J$ = 19.08). The location of this latter is 
coincident with the radio emission peak of the 1.4 GHz VLA image, 
so we consider J1329$-$3122a associated to the elliptical cluster galaxy; 

\parn $-$ J1330$-$3143b : this source is identified with the spiral \#10869, but
other two cluster galaxies fall within the radio isophotes of this extended 
source. 

\parn $-$ J1330$-$3127 : this identification is uncertain since the candidate 
optical counterpart, which falls within the radio contours, is misplaced with 
respect to the radio emission peak.  

\parn $-$ J1331$-$3139 : this source is the blend of the two J1331$-$3139a
and J1331$-$3139b radio sources of the ATCA--22cm catalogue. The latter is 
identified with the 17.97--mag quasar.

\parn $-$ J1331$-$3143 : this source is probably the blend of more sources. 
A galaxy with $b_J$ = 16.82 and z $=$ 0.044 is located within the radio
contours, at $\sim$ 45 arcsec from the radio emission peak.

\parn $-$ J1331$-$3155b: a spiral galaxy, with $b_J$ = 15.39 and v = 4505
km s$^{-1}$ is located at the edge of the radio contours, at $\sim$ 37 
arcsec from the emission peak.

\parn $-$ J1332$-$3141a: a cluster galaxy with $b_J$ = 16.96 and v = 10732
km s$^{-1}$ lies within the radio contours of the source, at
$\sim$ 50 arcsec from the emission peak.

\parn $-$ J1334$-$3132: this source is the blend of the three ATCA 22cm 
sources J1334$-$3132a, J1334$-$3132b e J1334$-$3132c. J1334$-$3132c 
is associated with the 17.30-mag elliptical galaxy \#13281. This galaxy 
falls within the radio contours of the 1.4 GHz VLA image, 
but it is dislocated with respect to the emission peak.

\end{document}